\documentclass[12pt,aps,pre,reprint,twocolumn,superscriptaddress]{revtex4-2}

\usepackage{graphicx,psfrag,amsmath,amssymb,amsfonts,bbm,latexsym,color,dcolumn,bm,mathbbol}
\usepackage[framemethod=tikz]{mdframed}
\usepackage[colorlinks=true,allcolors=blue]{hyperref}
\usepackage{soul} 
\usepackage{cancel}

\usepackage{array}
\usepackage{booktabs}
\usepackage{footnote}
\usepackage{threeparttable}

\usepackage{svg}

\usepackage{natbib}
\usepackage{amsmath,amssymb}             
\usepackage{color}

\usepackage{enumerate}

\definecolor{szlcolor}{rgb}{0.8, 0, 0.6}

\definecolor{bluecolor}{rgb}{0, 0, 1.0}

\definecolor{redcolor}{rgb}{1, 0, 0}

\usepackage[normalem]{ulem}
\definecolor{jfcolor}{rgb}{0.1, 0.0, 0.9}

\begin{document}

\title{Curvature-induced clustering of cell adhesion proteins}

\date{\today}

\author{Shao-Zhen Lin}
\affiliation{Aix Marseille Univ, Université de Toulon, CNRS, CPT (UMR 7332), Turing Centre for Living systems, Marseille, France}
\author{Jacques Prost}
\affiliation{Laboratoire Physico-Chimie Curie, UMR 168, Institut Curie, PSL Research University, CNRS, Sorbonne Université, 75005 Paris, France}
\affiliation{Mechanobiology Institute, National University of Singapore, 117411 Singapore}
\author{Jean-François Rupprecht}
\affiliation{Aix Marseille Univ, Université de Toulon, CNRS, CPT (UMR 7332), Turing Centre for Living systems, Marseille, France}

\begin{abstract} 
Cell adhesion proteins typically form stable clusters that anchor the cell membrane to its environment. Several works have suggested that cell membrane protein clusters can emerge from a local feedback between the membrane curvature and the density of proteins. Here, we investigate the effect of such curvature-sensing mechanism in the context of cell adhesion proteins. We show how clustering emerges in an intermediate range of adhesion and curvature-sensing strengths. We identify key differences with the tilt-induced gradient sensing mechanism we previously proposed (Lin et al., arXiv:2307.03670, 2023). 


\end{abstract}

\maketitle


\section{Introduction}

Cells rely on specific proteins to bind to their environment. In particular, cadherin mediate the attachment between cells, while integrins mediate the attachment between cells and the extracellular matrix. Cell adhesion proteins are known to play a central role in several processes critical in the development and maintenance of tissues and organs \cite{Gumbiner1996,Ladoux2012,Schwarz2013,Sun2019,Janiszewska2020}. 

Here we develop a generic, coarse-grained model for the supramolecular assembly of cell adhesion proteins. Our inspiration comes from recent experiments on spreading cells \cite{Changede2015, Yu2015, Changede2019}, which show that integrins form circular clusters at the cell leading edge. These clusters form within $3 \ \rm min$, which is short as compared to their lifetime, suggesting that these clusters are stable. 

To interpret the formation of such stable clusters, we proposed in Ref. \cite{Lin2023} a tilt-induced clustering mechanism in which gradients in the membrane height allow for the development of a mean tilt of cell adhesion proteins. Such mean tilt along membrane gradients relaxes the conformational energy of cell adhesion proteins. Cluster form when the gain in conformational energy exceeds the membrane deformation cost.

Here we consider an alternative mechanism for the formation of stable clusters in which cell adhesion proteins are sensitive to the local membrane curvature. A spontaneous membrane curvature generically emerges when the symmetry between the inner and outer leaflets of the cell membrane is broken \cite{Marcerou1984,doi:10.1080/000187399243428}. Such symmetry breaking can be caused by molecules or proteins that bind to a specific leaflet (e.g., crenator molecules \cite{Sheetz1974}, epsin \cite{Stachowiak2012}, or Bin-Amphiphysin-Rvs proteins \cite{McMahon2015, Ramakrishnan2013}), or, in the case of transmembrane proteins, due to a difference in the area occupied within each leaflets \cite{Zakany2020}. Here, we consider that the local concentration in cell adhesion proteins modulates the membrane spontaneous curvature. Our approach is agnostic of the specific microscopic mechanism involved in setting up such spontaneous curvature. 

We point out that the two mechanisms of gradient-sensing (described in Ref. \cite{Lin2023}) and  curvature-sensing (described here) are not mutually exclusive; on the contrary, they likely complement each other, as orders of magnitude suggest. Both mechanisms predict similar spatial patterns, e.g. stable circular clusters and line structures which are reminiscent of patterns observed in cells: nascent adhesion organize as disk-like clusters \cite{Changede2015, Yu2015, Changede2019} while focal or fibrillar adhesion are linear structures \cite{Zamir2000}. Despite these similarities, we identify a key differences, notably regarding the role of the adhesion strength on the emergence of clusters. 

Several works tackled the role of the membrane-to-ligand distance on the cell adhesion binding affinity and on the growth rate of clusters \cite{Bihr2012, Bihr2015}. However, the possibility that cell adhesion proteins could generate a spontaneous curvature appears unexplored - with the exception of a recent theoretical study \cite{Li2019}, which focuses on the different problem of evaluating of the protein binding affinity. 



\begin{figure}[t!]
\centering
\includegraphics[width=8.6cm]{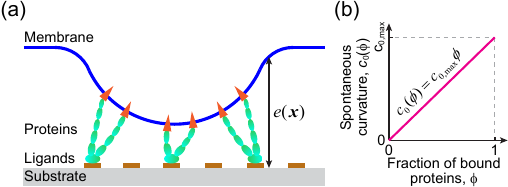}
\caption{\label{fig_ModelSketch} 
(a) Model sketch of the membrane--protein--substrate system. Proteins (green) pin the membrane (blue) to the substrate (grey) by adhesion to ligands (brown). Triangles (orange) indicate that the adhesion protein induces a local spontaneous curvature into the cell membrane. 
(b) The spontaneous curvature $c_{0}(\phi)$ of a cell membrane is assumed to be a linear function of the fraction of bound adhesion proteins $\phi$. 
}
\end{figure}

This article is organized as follows. 
In Sec. \ref{sec:Model}, we present a theoretical model to describe the membrane–protein–substrate system (see Fig. \ref{fig_ModelSketch}(a)) that describes the mixing of proteins, membrane deformation elasticity, and membrane--substrate adhesion. We consider a simulation protocol that allows us to reach a state that minimizes the free energy of the system. 
In Sec. \ref{sec:Results}, we first apply our theory and perform numerical simulations to investigate the cluster formation of cell adhesion proteins, focusing on the spontaneous curvature and cell adhesion parameters. We next present an analytical criteria for the stability of the homogeneous state, which we show accounts for the type of patterns observed in simulations. 
In Sec. \ref{sec:Discussion}, we interpret the transition from hexagonally-arranged circular clusters to lines through a mapping to the Swift--Hohenberg theory. We then discuss the relation of the curvature-sensing mechanism proposed here to our previously proposed gradient-sensing mechanism \cite{Lin2023}, as well as applications to the interpretation of experimental observations. 
Finally, we give our main conclusions in Sec. \ref{sec:Conclusion}.

\section{Model and simulation} \label{sec:Model}

\subsection{Theoretical model}

We consider a membrane--protein--substrate system, as shown in Fig. \ref{fig_ModelSketch}(a). 
Our model is defined in terms of two fields: (1) the fraction of bound cell adhesion proteins among other molecules, $\phi \left( \bm{x} \right) \in (0,1)$; (2) the height of the membrane with respect to the substrate, $e\left( \bm{x} \right)$. 
For a fully attached membrane ($\phi \rightarrow 1$), $e \rightarrow 0$; while for a fully detached membrane ($\phi \rightarrow 0$), $e \rightarrow e_0$ with $e_0$ standing for the membrane rest-length height, see Fig. \ref{fig_ModelSketch}(a). 

We consider the total free energy for the membrane--protein--substrate system in the following form: 
\begin{equation}
F[e,\phi] =  \int \mathrm{d}^2\bm{x} \left\{ f_{\rm FH}[\phi] + f_{\rm Hel}[e,\phi] + f_{\rm adh}[e,\phi] \right\} , \label{eq:FreeEnergy}
\end{equation}
where $f_{\rm FH}$, $f_{\rm Hel}$, and $f_{\rm adh}$ represent the free energy densities associated with protein mixing, membrane deformation, and membrane--substrate adhesion, respectively. 
We next explain these three terms in details:
\begin{itemize}
    \item Clustering comes at an entropy cost. As in Ref \cite{Lin2023}, we consider an entropy associated with protein binding in the form
\begin{equation}
f_{\rm FH} =  \frac{k_B T}{a} \left[ \phi\ln\phi + (1-\phi)\ln(1-\phi) \right] \\
+ \frac{D_{\phi}\left(\nabla\phi\right)^2}{2} , 
\end{equation}
where $k_B T$ is the thermal energy and $a$ is the inverse areal density of binders (as in Flory-Huggins theory \cite{Huggins1941,Flory1941}); $D_{\phi}$ is a gradient energy coefficient, which controls the width of the cluster interfaces \cite{Raote2020}. 
\item Adhesion molecules typically pin the cell membrane at a relatively short distance $\sim 30 \ \mathrm{nm}$, against  $\sim 110 \ \mathrm{nm}$ in regions where the cell membrane only interacts with the substrate due to glycocalyx steric interactions \cite{Bihr2012}. As in Ref. \cite{Lin2023}, we propose the following free energy for such adhesion-mediated interaction between the membrane and the flat substrate 
\begin{equation}
f_{\rm adh} = \frac{1}{2}{{k}_{0}}{{e}^{2}}-{{k}_{0}}{{e}_{0}}\left( 1-\phi  \right)e - h_{\phi}\phi, \label{eq:fadh} 
\end{equation}
where $k_0$ is the membrane--substrate binding elastic constant, $e_0$ is the height difference between the adhered state and the detached state, and $h_{\phi}$ is the chemical potential of protein--substrate binding. 
\item Membrane deformation generically comes at an energy cost. The spontaneous curvature $c_0$ characterizes the non-deformed, stress-free curvature of the membrane. Here, inspired by previous works \cite{Gov2018}, we consider a density-dependent spontaneous curvature $c_0(\phi)$, within the following Helfrich free energy, 
\begin{equation}
f_{\rm Hel} = \frac{1}{2}\sigma (\nabla e)^2 + \frac{1}{2}\kappa\left[ \nabla^2 e - c_0 (\phi) \right]^2 , \label{eq:HelfrichFreeEnergy}
\end{equation}
where $\sigma$ is the surface tension and $\kappa$ is the bending stiffness. As per the standard convention, the curvature $\nabla^2 e$ is positive when the membrane is deforming away from the cell cytoplasm, see Fig. \ref{fig_ModelSketch}(a). 
We here consider a relation between $\phi$ and $c_0$ at the simplest, linear order: 
\begin{equation}
c_0(\phi) = c_{0,\min} + (c_{0,\max} - c_{0,\min}) \phi, 
\label{eq:c0_phi_coupling}
\end{equation} 
where $c_{0,\max}$ (resp. $c_{0,\min}$) quantifies the maximal (resp. minimal) spontaneous curvature that can be achieved for a maximal (resp. minimal) fraction of adhesion proteins. 
However, substituting Eq. (\ref{eq:c0_phi_coupling}) into Eq. \eqref{eq:HelfrichFreeEnergy} and examining the total free energy, we find that the constant part $c_{0,\min}$ is equivalent to renormalizing the adhesion energy $h_{\phi} \to h_{\phi} - \kappa c_{0,\min} (c_{0,\max} - c_{0,\min})$. Therefore, without loss of generality, we assume $c_{0,\min} = 0$ for simplicity in our present study, see Fig. \ref{fig_ModelSketch}(b). 

\end{itemize}

\subsection{Numerical simulation}

To obtain the minimum energy state of the system, we consider the following annealing dynamics
\begin{align}
\frac{\partial e}{\partial t}&=-\frac{\text{ }\!\!\delta\!\!\text{ }F}{\text{ }\!\!\delta\!\!\text{ }e} + \eta \left( \bm{x},t \right) , \label{eq_dedt_0} \\ 
\frac{\partial \phi }{\partial t}&=-\frac{\text{ }\!\!\delta\!\!\text{ }F}{\text{ }\!\!\delta\!\!\text{ }\phi } , \label{eq_dphidt_0}
\end{align}
where $\eta \left( \bm{x},t \right)$ is the decaying noise source, implemented as a Gaussian white noise with zero mean ($\left< \eta \left( \bm{x},t \right) \right> = 0$) and variance $\left< \eta \left( \bm{x},t \right) \eta \left( \bm{x}',t' \right) \right> = \Lambda(t)^2 \delta \left(\bm{x} - \bm{x}' \right) \delta \left(t - t' \right)$ whose intensity $\Lambda(t)$ is a decreasing function of time (see Appendix \ref{sec:Simulation_scheme} for details). 

Substituting Eq. \eqref{eq:FreeEnergy} into Eqs. \eqref{eq_dedt_0} and \eqref{eq_dphidt_0}, we obtain 
the following evolution equations
\begin{align}
\frac{\partial e}{\partial t} = & -{{k}_{0}}e-{{k}_{0}}{{e}_{0}}\phi +\sigma {{\nabla }^{2}}e-\kappa {{\nabla }^{2}}{{\nabla }^{2}}e \notag \\
& +\kappa {{c}_{0,\max }}{{\nabla }^{2}}\phi + {{k}_{0}}{{e}_{0}}+\eta \left( \bm{x},t \right) , \label{eq_dedt} 
\end{align}
and
\begin{align}
\frac{\partial \phi }{\partial t} = & -{{k}_{0}}{{e}_{0}}e-\kappa c_{0,\max }^{2}\phi +\kappa {{c}_{0,\max }}{{\nabla }^{2}}e+{{D}_{\phi }}{{\nabla }^{2}}\phi \notag \\
& -\frac{{{k}_{B}}T}{a}\ln \left( \frac{\phi }{1-\phi } \right)+{{h}_{\phi }} . \label{eq_dphidt}
\end{align}

We point out that the total number of adhesion molecules is not conserved through our an energy minimization process, which represents a key difference with the previous models reviewed in \cite{Gov2018}.

\begin{table}[t!]
\centering
\caption{List of default parameter values} \label{table_s1}
\begin{threeparttable}
{\def\arraystretch{1.8}
\begin{tabular}{p{1.5cm}<{\centering} p{4cm}<{\centering} p{3cm}<{\centering}}
\toprule[1.0pt]
Parameter & Description & Value \\ \midrule[0.5pt]
$e_0$ & Membrane rest-length height & $80 \ \rm nm$ \cite{Smith2008,Bihr2012} \\ 
$k_B T$ & Thermal energy & $4 \times 10^{-21} \ \rm J$ \\ 
$k_0$ & Membrane-substrate adhesion stiffness & $2 \times 10^{-5} \ k_B T \cdot \rm nm^{-4}$ \cite{Bihr2012} \\ 
$\kappa$ & Membrane bending stiffness & $10 \ k_B T$ \cite{Bihr2012,Weikl2018,Steinkuhler2019,Raote2020} \\ 
$\sigma$ & Membrane surface tension & $0.005 \ k_B T \cdot \rm nm^{-2}$ \cite{Popescu2006,Kozlov2015,Raote2020} \\ 
$a$ & Inverse areal density of binders & $100 \ \rm nm^2$ \cite{Xu2016,Raote2020} \\ 
$D_{\phi}$ & Gradient energy coefficient & $1 \ k_B T$ \\
\bottomrule[1.0pt]
\end{tabular}
}
\end{threeparttable}
\end{table}

\subsection{Default set of parameters} \label{sec:Parameters}
Based on previously reported experimental measurements, we consider the following parameter values: a typical height difference $e_0 = 80 \ \rm nm$; a distance between binders $d = 10 \ \rm nm$ which results in $a = 100 \ \rm nm^2$ \cite{Changede2015}; a cell membrane tension $\sigma = 2 \times 10^{-5} \ {\rm J \cdot m^{-2}} \approx 0.005 \ k_B T \cdot \rm nm^{-2}$ \cite{Kozlov2015,Raote2020} and a membrane bending rigidity $\kappa = 10 \ k_B T$ \cite{Raote2020,Steinkuhler2019,Weikl2018}. 
The binding energy $k_0 e_0^2 a \sim 10 \ k_B T$ \cite{Bihr2012,Bihr2015} yields the following value of the membrane--substrate binding stiffness: $k_0 \sim 10 k_B T / (e_0^2 a) \sim 10^{-5} \ k_B T \cdot \rm nm^{-4}$; therefore, we find a value of the effective stiffness, $k_0 a \sim 10^{-3} \ k_B T \cdot \rm nm^{-2}$, which is consistent with one provided in Ref. \cite{Bihr2012} (the parameter $\lambda$ therein). Further, we consider $D_{\phi} = k_B T$ and $h_{\phi} = k_0 e_0^2 = 0.1 \ k_B T \cdot \rm nm^{-2}$. 

To our knowledge, there is as yet no work dealing with the specific contribution of cell adhesion proteins (such as integrins or cadherins) to the spontaneous curvature of a cell membrane. Here, we propose to set the maximal spontaneous curvature at $c_{0,\max} \sim 0.1 \ \rm nm^{-1}$, a value within the range reported for membranes that include epsin1 \cite{Stachowiak2012} or SNARE \cite{Stratton2016} complexes (see also Refs. \cite{Kluge2022, Lipowsky2015}). 

In our simulations, we normalize the parameters by the length scale $e_0$ and the energy scale $k_B T$. We set our default set of non-dimensional parameters as below: $\tilde{k}_0 = k_0 e_0^4 / (k_B T) = 1000$, $\tilde{\kappa} = \kappa / (k_B T) = 10$, $\tilde{\sigma} = \sigma e_0^2/ (k_B T) = 32$, $\tilde{a} = a / e_0^2 = 1 / 64$, $\tilde{D}_{\phi} = D_{\phi} / (k_B T) = 1$. Such a set of parameters corresponds to the values of Table \ref{table_s1} in dimensionalized units.

\begin{figure*}[t!]
\centering
\includegraphics[width=16.8cm]{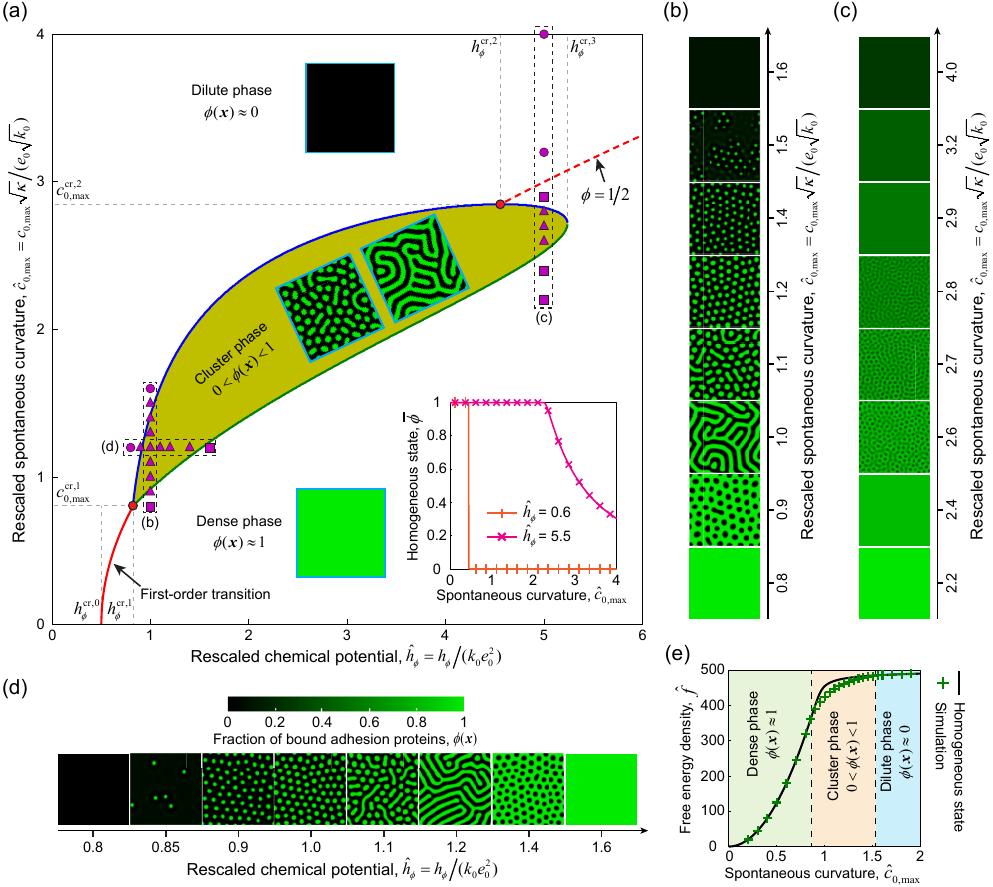}
\caption{\label{fig_PhaseDiagram} 
The chemical potential $h_{\phi}$ and the spontaneous curvature $c_{0,\max}$ dictate the cluster formation of curvature-inducing cell adhesion proteins. 
(a) Stability of the homogeneous state ($\bar{e}$, $\bar{\phi}$), regulated by the rescaled chemical potential $\hat{h}_{\phi} = h_{\phi} / (k_0 e_0^2)$ and the rescaled spontaneous curvature $\hat{c}_{0,\max} = c_{0,\max}\sqrt{\kappa} / (e_0 \sqrt{k_0})$. This phase diagram is obtained by linear stability analysis, where the yellow region indicates cluster formation ($\beta_{\min} < 0$, see Eq. (\ref{eq:betamin})). 
The red solid line refers to a first-order transition between the dense phase ($\phi \approx 1$) and the dilute phase ($\phi \approx 0$), see Eq. \eqref{eq:DenseToDiluteTransition}; while the red dashed line refers to the $\phi = 1/2$ line. See the inset for these two different behaviors. 
The green solid line (resp. the blue solid line) refers to the lower (resp. upper) bound (with respect to $c_{0,\max}$) of the cluster formation regime, obtained by numerical calculations of $\beta_{\min} = 0$. 
Purple symbols represent numerical simulation data (see b-d for the corresponding patterns): squares refer to the dense phase; triangles refer to the cluster phase; circles refer to the dilute phase. 
The two red circles indicate the triple point (bottom-left one) and the critical point (top-right one). 
The square box patterns refer to some typical patterns of these three different phases, obtained by numerical simulations. 
The gray dashed lines indicate critical values of $h_{\phi}^{\rm cr,0}$, $h_{\phi}^{\rm cr,1}$, $h_{\phi}^{\rm cr,2}$, $h_{\phi}^{\rm cr,3}$, $c_{0,\max}^{\rm cr,1}$ and $c_{0,\max}^{\rm cr,2}$. 
Inset: The homogeneous state $\bar{\phi}$ as a function of the rescaled curvature $\hat{c}_{0,\max}$. 
(b, c) Typical patterns of bound cell adhesion proteins at different values of $c_{0,\max}$. 
Parameters: (b) $\hat{h}_{\phi} = h_{\phi} / (k_0 e_0^2) = 1$; (c) $\hat{h}_{\phi} = h_{\phi} / (k_0 e_0^2) = 5$. 
(d) Typical patterns of bound cell adhesion proteins at different values of $h_{\phi}$. Here $\hat{c}_{0,\max} = c_{0,\max} \sqrt{\kappa}/ (e_0 \sqrt{k_0}) = 1.2$. 
In (b-d), the color codes correspond to $\phi(\bm{x})$; simulation box size $L \times L$ with $L = 16 e_0$. 
(e) The free energy density $f$ as a function of the spontaneous curvature $c_{0,\max}$, where $\hat{h}_{\phi} = h_{\phi} / (k_0 e_0^2) = 1$. 
Solid line: theoretically predicted homogeneous steady state, obtained by numerical resolution of Eq. \eqref{eq_HomogeneousSteadyStateEquation_1}. Symbols: numerical simulations. 
See Sec. \ref{sec:Parameters} and Table \ref{table_s1} for other parameter values. 
}
\end{figure*}

\section{Results} \label{sec:Results}

\subsection{Simulation results}

We carried out a series of simulations for different values of $h_{\phi}$ and $c_{0,\max}$, with all other parameters fixed at the default values in Table \ref{table_s1}. In the cases $h_{\phi} < 0$ and $c_{0,\max} < 0$, no clusters appear. We expect such lack of cluster formation for $c_{0,\max} < 0$ as $e_0 >0$ in our default set of parameters, which implies that adhesion favors a positive membrane curvature. 

We therefore focus on the first quadrant $h_{\phi} > 0$ and $c_{0,\max} > 0$, see Fig. \ref{fig_PhaseDiagram}. We report on the final steady state reached in our simulation upon increasing the spontaneous curvature (starting from $c_{0,\max} = 0$) in the following regimes of
\begin{enumerate}
    \item very low chemical potential $0 < h_{\phi} < h_{\phi}^{\rm cr,0} = k_0 e_0^2 / 2$: simulations converge to a spatially homogeneous and dilute ($\phi \approx 0$) state, whatever the value of  $c_{0,\max}$.
    \item low chemical potential, $h_{\phi}^{\rm cr,0} = k_0 e_0^2 / 2 < h_{\phi} < h_{\phi}^{\rm cr,1} \approx 0.82 k_0 e_0^2$: simulations first converge to the homogeneous and dense ($\phi \approx 1$) state; as $c_{0,\max}$ is increased beyond a critical value $c^{\rm cr}_{0,\max}(h_{\phi})$, simulations converge to the homogeneous and dilute ($\phi \approx 0$) state (see the inset of Fig. \ref{fig_PhaseDiagram}(a)). The critical line $c^{\rm cr}_{0,\max}(h_{\phi})$ exhibits a $\sqrt{h_{\phi}-h_{\phi}^{\rm cr,0}}$ scaling, Fig. \ref{fig_PhaseDiagram}(a) -- a behavior reminiscent of a first-order transition in thermodynamics.
    \item intermediate chemical potential, $h_{\phi}^{\rm cr,1} \approx 0.82 k_0 e_0^2 < h_{\phi} < h_{\phi}^{\rm cr,3} \approx 5.25 k_0 e_0^2$: simulations first converge to the homogeneous and dense ($\phi \approx 1$) state, but the steady state reaches then undergoes a series of transition in patterns as the spontaneous curvature is increased: from low-density circular holes at $\phi \approx 0$,  into long connected lines, and then to dense circular clusters  (with $\phi \approx 1$), see Fig. \ref{fig_PhaseDiagram}(b). The intensity of such modulation is reduced for large chemical potential, Fig. \ref{fig_PhaseDiagram}(c). 
    \item high chemical potential, $h_{\phi} > h_{\phi}^{\rm cr,3} \approx 5.25 k_0 e_0^2$: the steady state remains spatially homogeneous. The homogeneous state $\bar{\phi}$ transitions smoothly from a dense state ($\bar{\phi} \approx 1$) to a dilute state ($\bar{\phi} \approx 0$), as the spontaneous curvature is increased from small values (i.e., $c_{0,\max} \ll c_{0,\max}^{\rm cr}(h_{\phi})$, see Eq. \eqref{eq:DenseToDiluteTransition}) to large values (i.e., $c_{0,\max} \gg c_{0,\max}^{\rm cr}(h_{\phi})$).
\end{enumerate}

Conversely, increasing the chemical potential at a fixed spontaneous curvature yields a transition from a homogeneous dilute state to clusters, and then to a homogeneous dense state,  Fig. \ref{fig_PhaseDiagram}(d). 

In the next two subsections, we seek to understand the results of these numerical simulations using an analytical approach.

\begin{figure*}[t!]
\centering
\includegraphics[width=17.5cm]{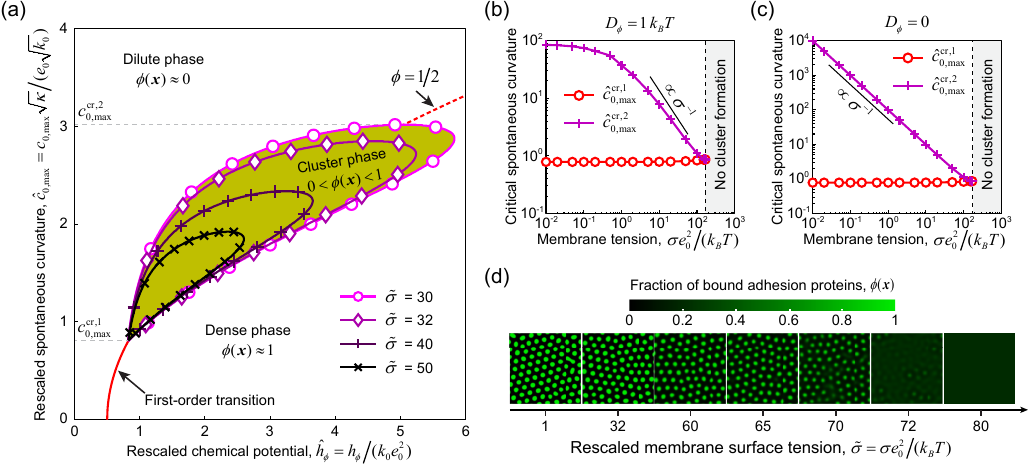}
\caption{\label{fig_TensionRegulation} 
The membrane surface tension $\sigma$ regulates cluster formation. 
(a) The phase diagram of cluster formation is regulated by the membrane surface tension $\tilde{\sigma} = \sigma e_0^2 / (k_B T)$. 
(b, c) The critical spontaneous curvature values $c_{0,\max}^{\rm cr,1}$ and $c_{0,\max}^{\rm cr,2}$ as a function of the membrane surface tension $\sigma$: (b) $D_{\phi} = 1 \ k_B T$; (c) $D_{\phi} = 0$. These data points are obtained from numerical resolution of the condition $\beta_{\min} = 0$, defined in Eq. (\ref{eq:betamin}).
(d) The membrane surface tension $\sigma$ regulates the cluster patterns of cell adhesion proteins; snapshots of the steady state reached in simulations for several values of the surface tension (parameter values: $\hat{h}_{\phi} = h_{\phi} / (k_0 e_0^2) = 1$, $\hat{c}_{0,\max} = c_{0,\max} \sqrt{\kappa} / (e_0 \sqrt{k_0}) = 1.2$; other parameter values in Sec. \ref{sec:Parameters} and Table \ref{table_s1}). 
}
\end{figure*}

\subsection{Homogeneous steady state} \label{sec:homogeneous_state}

\paragraph*{Theoretical solution of homogeneous states} Here we derive the expression of the spatially homogeneous solutions that minimize the total free energy. Solving for the condition $\delta F/\delta \phi = 0$, we find that such homogeneous steady state $(\bar{e},\bar{\phi})$ read 
\begin{equation}
\bar{e} = {{e}_{0}}\left( 1-\bar{\phi } \right), \quad \mathrm{and}  \quad  g\left( {\bar{\phi }} \right)  = 0, \label{eq_HomogeneousSteadyStateEquation_1}
\end{equation}
where
\begin{equation} \label{eq:gfunction1}
g\left( {\bar{\phi }} \right) = \left( \kappa c_{0,\max }^{2} - {{k}_{0}}e_{0}^{2} \right)\bar{\phi }+\frac{{{k}_{B}}T}{a}\ln \left( \frac{{\bar{\phi }}}{1-\bar{\phi }} \right)+{{k}_{0}}e_{0}^{2}-{{h}_{\phi }}. \notag 
\end{equation}
Given that $g\left( \bar{\phi }\to 0 \right)=-\infty$ and $g\left( \bar{\phi }\to 1 \right)=+\infty$, the equation $g(\phi) = 0$ has at least one root within the interval $(0,1)$, and the number of real roots will be odd. In practice, we observe either one or three roots. 
The transition from one to three roots is achieved at the specific point where $\min_{\phi \in (0,1)} \left\{ \frac{\mathrm{d}g}{\mathrm{d}\phi} \right\} = 0$. In Appendix \ref{sec:Analysis_h_phi_3}, we find that this point corresponds to 
\begin{equation}
h_{\phi}^{\rm cr,1} = k_0 e_0^2 - 2\frac{k_B T}{a} , \label{eq:hphi_cr_1_approximation}
\end{equation}
and
\begin{equation}
c_{0,\max}^{\rm cr,1} = \sqrt{\frac{k_0 e_0^2}{\kappa} - \frac{4 k_B T}{a\kappa}} . \label{eq:c0max_cr_1_approximation}
\end{equation}
In a region of parameters within the quadrant $h_{\phi} < h_{\phi}^{\rm cr,1}$ and $c_{0,\max} < c_{0,\max}^{\rm cr,1}$, three homogeneous states coexist; elsewhere, there is only one homogeneous state, see Fig. \ref{fig_NumberOfHomogeneousStates}.

\paragraph*{Free energy of homogeneous states} The free energy density of the spatially homogeneous state denoted $(\bar{e},\bar{\phi})$ reads
\begin{align}
f = & \ \frac{{{k}_{B}}T}{a}\left[ \bar{\phi} \ln \bar{\phi} +\left( 1-\bar{\phi}  \right)\ln \left( 1-\bar{\phi}  \right) \right] +\frac{1}{2}\kappa c_{0,\max }^{2}{{\bar{\phi} }^{2}} \notag \\ 
& -\frac{1}{2}{{k}_{0}}e_{0}^{2}{{\left( 1-\bar{\phi}  \right)}^{2}}-{{h}_{\phi }}\bar{\phi} . 
\end{align}
The later quantity increases with the value of the spontaneous curvature $c_{0,\max}$, see Fig. \ref{fig_PhaseDiagram}(e). 

In the fully detached state limit $\phi \approx 0$, the free energy density reads $f_{\rm attached} \approx -h_{\phi} + \kappa c_{0,\max}^2 / 2$. In the fully attached state limit $\phi \approx 1$, the free energy density reads $f_{\rm attached} \approx -h_{\phi} + \kappa c_{0,\max}^2 / 2$. The detached state is more stable than the attached one when $c_{0,\max} > c_{0,\max}^{\rm cr}$, 
with
\begin{equation}
c_{0,\max}^{\rm cr} = \sqrt{\frac{2 (h_{\phi} - h_{\phi}^{\rm cr,0})}{\kappa}} , \label{eq:DenseToDiluteTransition}
\end{equation}
where $h_{\phi}^{\rm cr,0} = k_0e_0^2/2$.

Equation (\ref{eq:DenseToDiluteTransition}) provides a quantitative prediction of the transition from the dense to dilute phases observed in our simulations, see the red line in Fig. \ref{fig_PhaseDiagram}(a). In particular, for $\hat{h}_\phi = h_{\phi} / (k_0 e_0^2) = 0.6$, Eq. (\ref{eq:DenseToDiluteTransition}) yields $\hat{c}_{0,\max} = c_{0,\max} \sqrt{\kappa} / (e_0 \sqrt{k_0}) \approx 0.447$, which agrees with the value $\hat{c}_{0,\max} = c_{0,\max} \sqrt{\kappa} / (e_0 \sqrt{k_0}) \approx 0.45$ observed in numerical simulations. 

\subsection{Linear stability analysis} \label{sec:LinearStability}

We next consider the stability of the homogeneous state $\left( \bar{e},\bar{\phi } \right)$. Around $\left( e,\phi \right)=\left( \bar{e},\bar{\phi } \right)$, the second-order variation of the free energy $F$ reads, 
\begin{equation}
{{\text{ }\!\!\delta\!\!\text{ }}^{2}}F = \frac{1}{{{\left( 2\text{ }\!\!\pi\!\!\text{ } \right)}^{2}}}\int{{{\text{d}}^{2}}\bm{q}\bar{\bm{\Phi}}\left( \bm{q} \right)\cdot \bm{J}\left( \bm{q} \right)\cdot \bm{\Phi}\left( \bm{q} \right)} , 
\end{equation}
where $\bm{\Phi}\left( \bm{q} \right) = ( \widetilde{\text{ }\!\!\delta\!\!\text{ }e}\left( \bm{q} \right) , \widetilde{\text{ }\!\!\delta\!\!\text{ }\phi }\left( \bm{q} \right) )$ and $\bar{\bm{\Phi}}(\bm{q}) = \bm{\Phi}(-\bm{q})$ is the conjugate complex of $\bm{\Phi}(\bm{q})$ with $\widetilde{\text{ }\!\!\delta\!\!\text{ }e}\left( \bm{q} \right)=\int{{{\text{d}}^{2}}\bm{q}\text{ }\!\!\delta\!\!\text{ }\hat{e}\left( \bm{x} \right)\exp \left( -\text{i}\bm{q}\cdot \bm{x} \right)}$ and $\widetilde{\text{ }\!\!\delta\!\!\text{ }\phi }\left( \bm{q} \right)=\int{{{\text{d}}^{2}}\bm{q}\text{ }\!\!\delta\!\!\text{ }\hat{\phi} \left( \bm{x} \right)\exp \left( -\text{i}\bm{q}\cdot \bm{x} \right)}$
being the Fourier transforms of non-dimensional perturbations $\delta \hat{e} =  {\delta e} / e_0$ and $\delta \hat{\phi} = {\delta \phi}$, respectively.
The Jacobian matrix $\bm{J}(\bm{q}) = [J_{ij}(\bm{q})]_{2 \times 2}$ reads 
\begin{equation}
\begin{split}
& J_{11} = {{k}_{0}}e_0^2 + \sigma e_0^2 {{\left| \bm{q} \right|}^{2}} + \kappa e_0^2 {{\left| \bm{q} \right|}^{4}} , \\
& J_{22} = \dfrac{{{k}_{B}}T}{a}\dfrac{1}{\bar{\phi }\left( 1-\bar{\phi } \right)}+\kappa c_{0,\max }^{2}+{{D}_{\phi }}{{\left| \bm{q} \right|}^{2}} , \\ 
& J_{12} = J_{21} = {{k}_{0}}{{e}_{0}^2} + \kappa e_0 {{c}_{0,\max }}{{\left| \bm{q} \right|}^{2}} . 
\end{split} \label{eq:JacobianMatrix}
\end{equation}
Since the matrix $\bm{J}(\bm{q})$ is symmetric, the eigenvalues of $\bm{J}\left( \bm{q} \right)$ are real; these read
\begin{equation}
{{\lambda }_{+}}=\frac{\alpha +\sqrt{{{\alpha }^{2}}-4\beta }}{2}\ ,\ {{\lambda }_{-}}=\frac{\alpha -\sqrt{{{\alpha }^{2}}-4\beta }}{2} \label{eq:Eigenvalues}
\end{equation}
with $\alpha = J_{11} + J_{22}$ and $\beta = J_{11} J_{22} - J_{12}^2$. Given Eq. \eqref{eq:JacobianMatrix}, we find that
\begin{equation}
\alpha (q) = \alpha_0 + \alpha_1 q^2 + \alpha_2 q^4 , 
\end{equation}
with the coefficients, 
\begin{equation}
\begin{split}
& {{\alpha }_{0}}=\frac{{{k}_{B}}T}{a}\frac{1}{\bar{\phi }\left( 1-\bar{\phi } \right)}+{{k}_{0}}e_{0}^{2}+\kappa c_{0,\max }^{2} \\ 
& {{\alpha }_{1}}=\sigma e_{0}^{2}+{{D}_{\phi }} \\ 
& {{\alpha }_{2}}=\kappa e_{0}^{2}
\end{split}
\end{equation}
Similarly, we find that
\begin{equation}
\beta \left( q \right)={{\beta }_{0}}+{{\beta }_{1}}{{q}^{2}}+{{\beta }_{2}}{{q}^{4}}+{{\beta }_{3}}{{q}^{6}} , 
\end{equation}
with the coefficients, 
\begin{equation}
\begin{split}
{{\beta }_{0}} = & \ \frac{{{k}_{0}e_0^2}{{k}_{B}}T}{a}\frac{1}{\bar{\phi }\left( 1-\bar{\phi } \right)} + {{k}_{0}e_0^2}\kappa c_{0,\max }^{2} - k_{0}^{2}e_{0}^{4} , \\ 
{{\beta }_{1}} = & \ \frac{\sigma e_0^2 {{k}_{B}}T}{a}\frac{1}{\bar{\phi }\left( 1 - \bar{\phi } \right)} + {{k}_{0}e_0^2}{{D}_{\phi }} + \sigma e_0^2 \kappa c_{0,\max }^{2} \\ 
& -2{{k}_{0}}{{e}_{0}^3} \kappa {{c}_{0,\max }} , \\ 
{{\beta }_{2}} = & \ \frac{\kappa e_0^2 {{k}_{B}}T}{a}\frac{1}{\bar{\phi }\left( 1-\bar{\phi } \right)} + \sigma e_0^2 {{D}_{\phi }} , \\ 
{{\beta }_{3}} = & \ \kappa e_0^2 {{D}_{\phi }} , 
\end{split} \label{eq:beta_i_expression}
\end{equation}
where $\bar{\phi}$ is given by Eq. \eqref{eq_HomogeneousSteadyStateEquation_1}. 

Note that $\alpha > 0$ and ${{\alpha }^{2}}-4\beta >0$ hold for arbitrary parameters and all wavenumber $q$, indicating $\lambda_{+} > 0$ for all $q$. Thus the stability condition of the homogeneous state $\left( e,\phi  \right)=\left( \bar{e},\bar{\phi } \right)$ is solely imposed by the sign of $\lambda_{-}(q)$. However, given that $\alpha (q) > 0$ for all $q$ (since $\alpha_0 > 0$, $\alpha_1 > 0$ and $\alpha_2 >0$), we find that the stability condition that $\lambda_{-}(q)>0$ for all $q$ is in fact equivalent to the condition that $\beta (q) > 0$ for all $q$, hence to 
\begin{equation}
\beta_{\min}\triangleq \underset{q}{\mathop{\min }}\,\left[ \beta \left( q \right) \right] > 0. \label{eq:betamin}
\end{equation}
After some algebra, we derive the following expression
\begin{equation}
\beta_{\min} = \left\{ \begin{aligned}
& \ \ \ \ \ \ \ \ \ \ \ \ \ {{\beta }_{0}}\ \ \ \ \ \ \ \ \ \ \ \ \ \ \ \ \ \ , \ \ \ {{\beta }_{1}}\ge 0 \\ 
& {{\beta }_{0}}+{{\beta }_{1}}q_{s}^{2}+{{\beta }_{2}}q_{s}^{4}+{{\beta }_{3}}q_{s}^{6} \ \ \ , \ \ \ {{\beta }_{1}}<0 \\ 
\end{aligned} \right.
\label{eq:betaminanalytical}
\end{equation}
where 
\begin{equation}
q_s = \sqrt{\frac{-{{\beta }_{2}}+\sqrt{\beta _{2}^{2}-3{{\beta }_{1}}{{\beta }_{3}}}}{3{{\beta }_{3}}}} . \label{eq:qs}
\end{equation}
When $\beta_{\min} > 0$ (resp. $\beta_{\min} < 0$), the homogeneous steady state ($\bar{e}$, $\bar{\phi}$) is stable (resp. unstable). We note that $\beta_2 > 0$ and $\beta_3 > 0$, such that at large wavenumbers $q \to +\infty$, $\beta(q) \to +\infty$, suggesting the system is always stable at small scales. 

\paragraph*{Stability of homogeneous states for large $h_{\phi}$ and $c_{0,\max}$} Inspecting Eqs. \eqref{eq:beta_i_expression} and (\ref{eq:betaminanalytical}) in the large spontaneous curvature limit, we find that the condition 
\begin{align}
c_{0,\max} > \max \left\{ \frac{2 k_0 e_0}{\sigma} , e_0\sqrt{\frac{k_0}{\kappa}} \right\} , \label{eq:c0max_cr2_upper_bound}
\end{align}
ensures that $\beta_0 > 0$ and $\beta_1 > 0$, hence that $\beta_{\min} > 0$ and that the homogeneous steady state is stable. Similarly, in the large chemical potential limit, e.g., 
\begin{align}
h_{\phi} \gg \kappa c_{0,\max}^2 + \frac{k_B T}{a} , 
\label{eq:h0max_upper_bound}
\end{align}
we also find that $\beta_0 > 0$ and $\beta_1 > 0$, hence that $\beta_{\min} > 0$. These two limit cases indicate that, in the ($h_{\phi}$, $c_{0,\max}$) parameter space, the regime of $\beta_{\min} < 0$ is a bounded region. 

\paragraph*{Phase diagram} We next systematically computed the quantity $\beta_{\min}$ defined in Eq. (\ref{eq:betaminanalytical}) in the ($h_{\phi}$, $c_{0,\max}$) parameter space. As expected from Eqs. \eqref{eq:c0max_cr2_upper_bound} and (\ref{eq:h0max_upper_bound}), we find that the condition $\beta_{\min} < 0$ can only be met within a bounded region of the parameter space, see Fig. \ref{fig_PhaseDiagram}(a), with $c_{0,\max}^{\rm cr,2} \approx 2.85 e_0 \sqrt{k_0 / \kappa}$ and $h_{\phi}^{\rm cr,3} \approx 5.24 k_0 e_0^2$ being the maximal values such that $\beta_{\min} < 0$  (see Fig. \ref{fig_PhaseDiagram}(a)).  More precisely, the region $\beta_{\min} < 0$ takes the shape of a leaf, see Fig. \ref{fig_PhaseDiagram}; by analogy to thermodynamics, we call the \textit{triple point} the top end of the leaf (i.e. the bottom-left red point in Fig. \ref{fig_PhaseDiagram}(a)), and the critical point the bottom end (i.e. the top-right red point in Fig. \ref{fig_PhaseDiagram}(a)). 

\paragraph*{Analytical approximation of the triple point} We find that the point of transition from one to three homogeneous states, defined in Sec. \ref{sec:homogeneous_state}, matches well the triple point ($h_{\phi}^{\rm cr,1}$, $c_{0,\max}^{\rm cr,1}$) defined in Fig. \ref{fig_PhaseDiagram}. The term triple point is chosen by analogy to the point of coexistence of the solid/liquid/gas phases.  With our default parameter set (see Sec. \ref{sec:Parameters}), Eqs. \eqref{eq:hphi_cr_1_approximation} and \eqref{eq:c0max_cr_1_approximation} yield $h_{\phi}^{\rm cr,1} \approx 0.87 k_0 e_0^2$ and $c_{0,\max}^{\rm cr,1} \approx 0.85 e_0 \sqrt{k_0/\kappa}$, both of which are close to those obtained by numerical resolution of the condition $\beta_{\min}= 0$, which read $h_{\phi}^{\rm cr,1} \approx 0.82 k_0 e_0^2$ and $c_{0,\max}^{\rm cr,1} \approx 0.80 e_0 \sqrt{k_0/\kappa}$. 

\paragraph*{Analytical approximation of the critical point} Here we derive an approximate expression of the point ($h_{\phi}^{\rm cr,2}$,$c_{0,\max}^{\rm cr,2}$) that we call \textit{critical}, by analogy to the point separating the supercritical fluid from the gas and liquid phases in standard thermodynamics theory. We consider the condition $\beta_{\min} = 0$ together with $\bar{\phi} = 1/2$ (such that $h_{\phi}$ and $c_{0,\max}$ satisfy Eq. \eqref{eq:DenseToDiluteTransition}, see Appendix \ref{sec:Analysis_h_phi_3} for details). In the limit $D_{\phi} \to 0$ and $\sigma \to 0$, we find that $c_{0,\max}^{\rm cr,2}$ and $h_{\phi}^{\rm cr,2}$ read
\begin{equation}
c_{0,\max }^{\text{cr},2}\approx \frac{2{{k}_{0}}{{e}_{0}}-4\sqrt{\dfrac{{{k}_{0}}{{k}_{B}}T}{a}}}{\sigma }, \label{eq:c0max_cr_2_approximation}
\end{equation}
and
\begin{equation}
h_{\phi }^{\text{cr},2}\approx \frac{1}{2}\left[ 1+\frac{4\kappa {{k}_{0}}}{{{\sigma }^{2}}}{{\left( 1-2\sqrt{\frac{{{k}_{B}}T}{a{{k}_{0}}e_{0}^{2}}} \right)}^{2}} \right]{{k}_{0}}e_{0}^{2}. \label{eq:hphi_cr_2_approximation}
\end{equation}

With our default parameter set (see Sec. \ref{sec:Parameters}), Eqs. \eqref{eq:c0max_cr_2_approximation} and \eqref{eq:hphi_cr_2_approximation} yield the critical values $h_{\phi}^{\rm cr,2} \approx 5.27 k_0 e_0^2$ and $c_{0,\max}^{\rm cr,2} \approx 3.09 e_0 \sqrt{k_0/\kappa}$. These are close to those obtained by numerical resolution of the condition $\beta_{\min}= 0$, which read $h_{\phi}^{\rm cr,2} \approx 5.40 k_0 e_0^2$ and $c_{0,\max}^{\rm cr,2} \approx 3.13 e_0 \sqrt{k_0/\kappa}$ for $D_{\phi} = 0$; and $h_{\phi}^{\rm cr,2} \approx 4.56 k_0 e_0^2$ and $c_{0,\max}^{\rm cr,2} \approx 2.85 e_0 \sqrt{k_0/\kappa}$ for $D_{\phi} = 1 \ k_B T$.

\begin{figure}[t!]
\centering
\includegraphics[width=8.6cm]{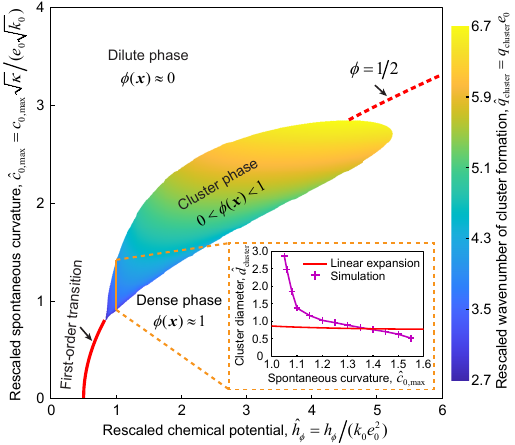}
\caption{\label{fig_ClusterSize} Phase diagram of the rescaled wavenumber $\hat{q}_{\rm cluster} = q_{\rm cluster} e_0$, with $q_{\rm cluster}$ defined in Eq. (\ref{eq:argminlambdaq}), as the wavenumber maximizing the instability growth rate $\lambda_{-}(q)$) as a function of the rescaled chemical potential $\hat{h}_{\phi} = h_{\phi} / (k_0 e_0^2)$ and the rescaled spontaneous curvature $\hat{c}_{0,\max} = c_{0,\max}\sqrt{\kappa} / (e_0 \sqrt{k_0})$.
Inset: The cluster diameter $d_{\rm cluster}$ measured in simulation (magenta crosses) and predicted cluster size (red line) as a function of the spontaneous curvature $c_{0,\max}$, where $\hat{h}_{\phi} = h_{\phi} / (k_0 e_0^2) = 1$. 
See Sec. \ref{sec:Parameters} and Table \ref{table_s1} for other parameter values. 
}
\end{figure}

\paragraph*{Impact of surface tension} In the limit of a vanishing cell membrane tension $\sigma \to 0$, we predict that $h_{\phi}^{\rm cr,2} \to +\infty$ and $c_{0,\max}^{\rm cr,2} \to +\infty$, based on Eqs. \eqref{eq:c0max_cr_2_approximation} and \eqref{eq:hphi_cr_2_approximation}. Conversely, we expect that a sufficiently large surface tension is sufficient can remove the possibility of clusters. 
Indeed, the triple point ($h_{\phi}^{\rm cr,1}$, $c_{0,\max}^{\rm cr,1}$) is insensitive to the membrane surface tension $\sigma$, as predicted by Eqs. (\ref{eq:hphi_cr_1_approximation}) and (\ref{eq:c0max_cr_1_approximation}), see Figs. \ref{fig_TensionRegulation}(a-c). By systematically estimating $\beta_{\min}$ for decreasing values of the surface tension $\sigma$, we found a regime where clusters could no longer be observed, see Figs. \ref{fig_TensionRegulation}(a-c). Indeed, starting from the default parameter set, the critical point ($h_{\phi}^{\rm cr,2}$, $c_{0,\max}^{\rm cr,2}$) shifts toward the triple point ($h_{\phi}^{\rm cr,1}$, $c_{0,\max}^{\rm cr,1}$) as the membrane surface tension $\sigma$ increases (Fig. \ref{fig_TensionRegulation}); we verify the $c_{0,\max}^{\rm cr,2} \sim \sigma^{-1}$ scaling predicted in Eq. \eqref{eq:c0max_cr_2_approximation}, see Fig. \ref{fig_TensionRegulation}(c). Thus increasing the cell membrane surface tension $\sigma$
tends to suppress cell adhesion protein clusters. 

\paragraph*{Comparison to numerical simulations} The instability of the homogeneous state criteria, $\beta_{\min} < 0$, is a good predictor for the formation of stable clusters in the numerical solution to Eqs. (\ref{eq_dedt_0}--\ref{eq_dphidt_0}), see Fig. \ref{fig_PhaseDiagram}. Clusters were still observed in a restricted range of parameters where $\beta_{\min} > 0$ suggesting that, within this regime, the homogeneous state is not the global minimum of the free energy. 

\paragraph*{Cluster size} We provide an analytical expression for the size of clusters observed in simulations. We call linear expansion length the quantity $d = 2 \pi/q_{\rm cluster}$, where $q_{\rm cluster}$ is the wavenumber at which the growth rate $\lambda_{-}(q) < 0$ (as defined by Eq. (\ref{eq:Eigenvalues})) is minimum:
\begin{align} \label{eq:argminlambdaq}
q_{\rm cluster} = \underset{q}{\mathrm{argmin}}(\lambda_{-}(q))
\end{align}
In simulations, the size of clusters is estimated through the quantity $d_{\rm cluster} = 2 \sqrt{A_{\rm cluster} / \pi}$, where $A_{\rm cluster}$ is the mean area of connected domains in which $\phi > 0.5$. The linear expansion length matches numerical simulations, in the regime of disk-like clusters (i.e. for $\hat{c}_{0,\max} = c_{0,\max}\sqrt{\kappa}/(e_0 \sqrt{k_0}) > 1.1$ in Fig. \ref{fig_ClusterSize} inset, and Fig. \ref{fig_PhaseDiagram}(b)), while larger deviations occur at low spontaneous curvatures, in the regime where lines form  (i.e. for $\hat{c}_{0,\max} = c_{0,\max}\sqrt{\kappa}/(e_0 \sqrt{k_0}) < 1.1$ in Fig. \ref{fig_ClusterSize} inset, Fig. \ref{fig_PhaseDiagram}(b)); however such deviations are to be expected given the definition of the cluster size considered here, which leads to a sharp increase of $d_{\rm cluster}$ when lines form.

\begin{figure}[t!]
\centering
\includegraphics[width=8.6cm]{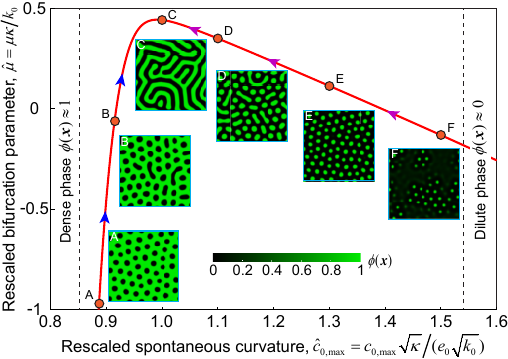}
\caption{\label{fig_SH_theory} 
Swift--Hohenberg theory interpretation of the clustering pattern transition of cell adhesion proteins upon varying the spontaneous curvature $c_{0,\max}$: line structures appear when the bifurcation parameter $\mu$ (see Eq. \eqref{eq:mu}) is maximal, see point C. 
The magenta and blue arrows indicate the two pattern transition routes discussed in the main text. 
Inset: Typical patterns of bound adhesion proteins. 
Parameters: $\hat{h}_{\phi} = h_{\phi} / (k_0 e_0^2) = 1$; see Sec. \ref{sec:Parameters} and Table \ref{table_s1} for other parameter values. 
}
\end{figure}

\section{Discussion} \label{sec:Discussion}

\subsection{Link to Swift--Hohenberg theory}

Our numerical simulations show the existence of circular clustering of hexagon-like patterns and line-structured patterns, see Fig. \ref{fig_PhaseDiagram}. This is reminiscent of the pattern formation in a Swift--Hohenberg theory \cite{Swift-Hohenberg1977,Hoyle2006}. Imposing the condition $e = e_0 (1 - \phi)$, Eq. \eqref{eq_dphidt} can be recast as the following Swift--Hohenberg equation (rescaling time in units of $\kappa e^2_0$; see Appendix \ref{sec:Link_to_SH_theory} for derivation details), 
\begin{equation}
\frac{\partial \delta\phi}{\partial t} = \left [ \mu - \left( q_c^2 + \nabla^2 \right)^2 \right] \delta\phi + m \left( \delta\phi \right) , \label{eq:Swift-Hohenberg}
\end{equation}
where $\delta\phi = \phi - \bar{\phi}$ is the perturbation away from the minimal energy homogeneous steady state; the critical wave-length reads
\begin{equation}
q_{c}^{2} = \frac{c_{0,\max} - c_{0,\max}^{\rm th}}{e_0} , \label{eq:qc_SH_theory}
\end{equation}
the bifurcation parameter is
\begin{equation}
\mu = \frac{{{k}_{0}}}{\kappa } - \frac{k_B T}{\kappa e_0^2 a \bar{\phi} (1-\bar{\phi})} + \frac{c_{0,\max }^{\text{th}} (c_{0,\max }^{\text{th}} - 2 c_{0,\max })}{e_{0}^{2}} , \label{eq:mu}
\end{equation}
and the non-linear function is
\begin{align}
m(\delta\phi) = \ & \frac{{{k}_{B}}T}{2 \kappa e_0^2 a } \frac{(1-2 \bar{\phi})}{\bar{\phi}^2(1-\bar{\phi})^2 } (\delta \phi)^2 \notag \\
& - \frac{{{k}_{B}}T}{3 \kappa e_0^2 a } \frac{1 - 3 \bar{\phi} (1-\bar{\phi})}{\bar{\phi}^3 (1-\bar{\phi})^3} (\delta \phi)^3 , \label{eq:nonlinear}
\end{align}
with $c_{0,\max}^{\rm th} = {(D_{\phi} + \sigma e_0^2)} / {(2 \kappa e_0)}$. 

In the Swift--Hohenberg theory, the wave-length $q_c$ controls the pattern size, while the bifurcation parameter $\mu$ controls the transition from a homogeneous state to spatial patterns; in particular, when a quadratic term is present in the non-linear function $m$, increasing $\mu$ first can yield a transition from a homoegeneous state to hexagonally arranged dots, and then to lines \cite{Hoyle2006}. 

Here, inspection of Eqs. (\ref{eq:qc_SH_theory}), (\ref{eq:mu}) and (\ref{eq:nonlinear}) shows that 
\begin{itemize}
    \item the non-linear term $m(\delta\phi)$ defined in Eq. (\ref{eq:nonlinear}) contains a quadratic contribution in $\delta\phi$ except when $\bar{\phi} = 1/2$ (i.e. $\bar{\phi}\rightarrow 1/2$ only in the high temperature regime),
    \item the bifurcation parameter $\mu$ defined in Eq. (\ref{eq:mu}) exhibits a maximum as a function of the spontaneous curvature $c_{0,\max}$, see Fig. \ref{fig_SH_theory}. This means that $\mu$ increases upon decreasing $c_{0,\max}$ from large values (dilute phase); this accounts well for the transition from hexagonally-arranged, circular clusters (dense phase) to dense lines observed in our numerical simulations upon decreasing $c_{0,\max}$, see magenta arrows in Fig. \ref{fig_SH_theory}.  Conversely, $\mu$ increases upon increasing $c_{0,\max}$ from small values (dense phase); this accounts well for the transition from hexagonally-arranged, circular clusters (dilute phase, or holes) to dilute lines (elongated holes) observed in our numerical simulations upon increasing $c_{0,\max}$, see blue arrows in Fig. \ref{fig_SH_theory}. 
    \item the wave-length $q_c$ defined in Eq. (\ref{eq:qc_SH_theory}) increases with $c_{0,\max}$; this accounts well for the decreasing cluster sizes $d_{\rm cluster}$ observed in our numerical simulations for increasing $c_{0,\max}$ values, see Fig. \ref{fig_ClusterSize}. 
\end{itemize}

\begin{figure}[t!]
\centering
\includegraphics[width=8.6cm]{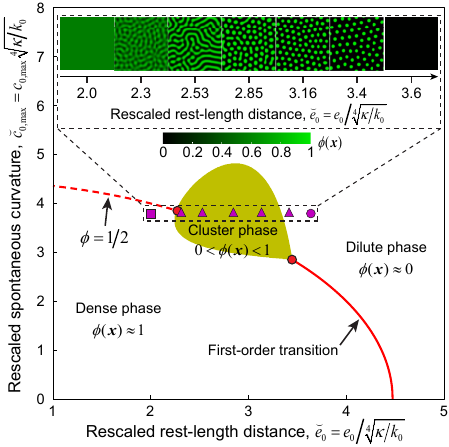}
\caption{\label{fig_e0_regulation} 
The membrane--substrate rest-length distance $e_0$ affects the cluster formation of cell adhesion proteins. 
Stability of the homogeneous state ($\bar{e}$, $\bar{\phi}$), regulated by the rescaled membrane--substrate rest-length distance $\check{e}_{0} = e_{0} / \sqrt[4]{\kappa/k_0}$ and the rescaled spontaneous curvature $\check{c}_{0,\max} = c_{0,\max} \sqrt[4]{\kappa/k_0}$. This phase diagram is obtained by linear stability analysis, where the yellow region corresponds to $\beta_{\min} < 0$. 
Purple symbols represent the location of the numerical simulation data represented in the inset: squares refer to the homogeneous dense state; triangles refer to the cluster state; circles refer to the homogeneous dilute state. 
The two red circles indicate the location of the triple and critical points. 
(Inset) Typical patterns of bound cell adhesion proteins at different values of $e_{0}$ where $\check{c}_{0,\max} = c_{0,\max} \sqrt[4]{\kappa/k_0} = 3.8$. 
See Sec. \ref{sec:Parameters} and Table \ref{table_s1} for other parameter values. 
}
\end{figure}

\subsection{Relation to the gradient-sensing mechanism \cite{Lin2023}}
In Ref. \cite{Lin2023}, we investigated the intrinsic tilt effect of cell adhesion proteins with respect to the membrane by introducing the free energy 
\begin{equation}
F_{\rm tilt} = \int \mathrm{d}^2\bm{x} \left[ -\frac{1}{2}\sigma_a \phi (\nabla e)^2 \right] , 
\end{equation}
with $\sigma_a > 0$.
We found that such a tilt energy contributes to an effective negative surface tension and leads to cluster formation. 

In a certain regime of cluster formation, the two parameters $\sigma_a$ and $c_{0,\max}$ identify each other; this can be well seen in the framework of the Swift--Hohenberg theory framework. 
However, unlike the tilt parameter $\sigma_a$, the spontaneous curvature parameter $c_{0,\max}$ also modulates the energy level of the adhered state.
This is why clusters disappear (by detaching) when $c_{0,\max}$ becomes too large, whereas clusters do not disappear even at large $\sigma_a$. 

In particular, in the absence of adhesion energy ($e_0 = 0$ or $k_0 = 0$, and $h_{\phi} = 0$), a sufficiently large tilt $\sigma_a > \sigma$ destabilizes the homogeneous state  \cite{Lin2023}. In contrast, the homogeneous state is always stable in the curvature-sensing mechanism considered here. 
Indeed, when $k_0 \to 0$, the coefficients $\beta_i$ simplify to $\beta_0 \approx 0$, $\beta_1 \approx \sigma e_0^2 k_B T / [a \bar{\phi} (1-\bar{\phi})] + \sigma e_0^2 \kappa c_{0,\max}^2 > 0$, $\beta_2 > 0$ and $\beta_3 > 0$.
Similarly, when $e_0 \to 0$ (vanishing attached/detached height difference), $\beta_0 \approx k_0 e_0^2 \{ k_B T / [a \bar{\phi} (1-\bar{\phi})] + \kappa c_{0,\max}^2 \} > 0$, $\beta_1 \approx e_0^2 \{ \sigma k_B T / [a \bar{\phi} (1-\bar{\phi})] +k_0 D_{\phi} + \sigma \kappa c_{0,\max}^2 \} > 0$, $\beta_2 > 0$, and $\beta_3 > 0$, hence the homogeneous steady state is stable; such behavior is observed in our systematic analysis on the role of $e_0$, see Fig. \ref{fig_e0_regulation}. 



\subsection{Experimental relevance}


Given the default parameter set defined in Table \ref{table_s1}, we find that the spontaneous curvatures of the triple and critical points are in the order of $c_{0,\max} \sim 0.1 \ \rm nm^{-1}$. 
As discussed in Sec. \ref{sec:Model}, such value lies within the range previously reported for the cell membrane \cite{Stachowiak2012,Stratton2016,Kluge2022, Lipowsky2015}. 

The transition from nascent adhesions (circular clusters) into focal adhesions (line structures) is associated with an increase in the strength of actin fibers, which pull on the adhesion sites \cite{Shemesh2005, Gov2006}. 
We propose that the strengthening effect of actin fibers on adhesion sites could be encompassed by a decrease in the height difference $e_0$ between the adhered state and the detached state. 
Starting from a condition of circular clusters (modeling nascent adhesions), we find that a decrease in the height difference $e_0$ leads to lines (modeling focal adhesions), see Fig. \ref{fig_e0_regulation}. 

We also find that increasing the membrane surface tension $\sigma$ leads to the disappearance of clustering patterns, either in favor of the homogeneous dilute state (for large spontaneous curvature)
or to the homogeneous dense state (for small spontaneous curvature). Such a transition is echoed by experimental observations that show the role of membrane surface tension on the assembly of cadherin aggregates \cite{Delanoe2004} or the disassembly of integrin-based fibrillar adhesion \cite{Zamir2000}. 


\section{Conclusion} \label{sec:Conclusion}

In conclusion, we have proposed a theoretical framework that accounts for a curvature-sensing mechanism to investigate the cluster formation phenomenon of cell adhesion proteins, e.g., integrins. 
Through theoretical analysis and numerical simulations, we show that coordinated cell adhesion and spontaneous curvature can lead to clustering patterns. 
Our simulations reveal various patterns of clusters, including hexagonal-arranged circular dots, long-curved stripe structures, and Turing-like patterns. 
We further show that these pattern transitions can be interpreted by the Swift--Hohenberg theory. 
We expect our findings to be useful in the near future to interpret experimental results on the clustering of integrins under various types of perturbations of the cell membrane properties.

\section*{Acknowledgements} 
J.-F. R. is hosted at the Laboratoire Adhésion Inflammation (LAI). The project leading to this publication has received funding from France 2030, the French Government program managed by the French National Research Agency (ANR-16-CONV-0001) and from Excellence Initiative of Aix-Marseille University - A*MIDEX. J.-F. R. is also funded by ANR-20-CE30-0023 COVFEFE.

\appendix

\section{Homogeneous steady state} \label{sec:Analytical_homogeneous_state}

\begin{figure}[t!]
\centering
\includegraphics[width=8.6cm]{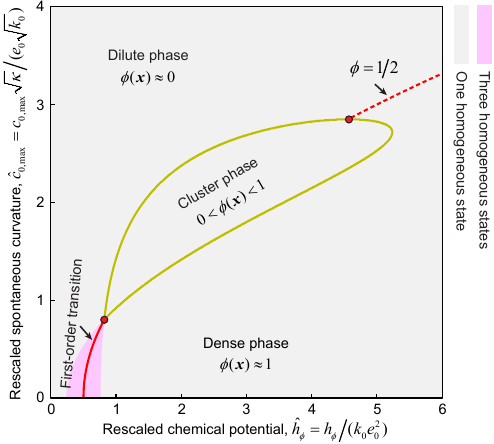}
\caption{\label{fig_NumberOfHomogeneousStates} 
Phase diagram of the number of homogeneous states, regulated by the rescaled chemical potential $\hat{h}_{\phi} = h_{\phi} / (k_0 e_0^2)$ and the rescaled spontaneous curvature $\hat{c}_{0,\max} = c_{0,\max}\sqrt{\kappa} / (e_0 \sqrt{k_0})$. 
This phase diagram is obtained by the numerical solution of Eq. \eqref{eq_HomogeneousSteadyStateEquation_1}. 
The region surrounded by the yellow line refers to the cluster phase ($\beta_{\min} < 0$), obtained by linear stability analysis. 
See Sec. \ref{sec:Parameters} and Table \ref{table_s1} for other parameter values. 
}
\end{figure}

\paragraph*{Uniqueness condition} A necessary condition to have only one solution to the condition $g=0$ is that the derivative of $g$ remains positive. Such derivative reads
\begin{equation}
\frac{\mathrm{d}g}{\mathrm{d}\phi} = \kappa c_{0,\max}^2-k_0 e_0^2 + \frac{k_B T}{a} \frac{1}{\phi(1-\phi)} . 
\end{equation}
whose minimum in $\phi$ reads
\begin{equation}
\min_{\phi \in (0,1)} \left\{ \frac{\mathrm{d}g}{\mathrm{d}\phi} \right\} = \kappa c_{0,\max}^2-k_0 e_0^2 + 4\frac{k_B T}{a} . \label{eq:min_dg_dphi}
\end{equation}
The later quantity remains positive under the condition
\begin{equation}
\left| c_{0,\max} \right| > \sqrt{\frac{k_0 e_0^2}{\kappa} - \frac{4 k_B T}{a\kappa}} . 
\end{equation}
Under such condition, there is one and only one homogeneous state, regardless of the value of $h_{\phi}$, see Fig. \ref{fig_NumberOfHomogeneousStates}. 

\paragraph*{Dilute and dense limits} We further analyze the behavior of solutions to the equation $g(\phi) = 0$ in the following dilute and dense limits.
\begin{itemize}
    \item 
In the dilute phase, i.e., $\bar{\phi }\to 0$, (in which $\bar{e}\to {{e}_{0}}$), we have 
\begin{equation}
g\left( \bar{\phi }\to 0 \right) \simeq \frac{{{k}_{B}}T}{a}\ln \bar{\phi}+{{k}_{0}}e_{0}^{2}-{{h}_{\phi }} . 
\end{equation}
In such dilute phase limit, the condition $g(\bar{\phi }) = 0$ leads to the following expression
\begin{equation}
\bar{\phi } = \exp \left[ \frac{a\left( {{h}_{\phi }}-{{k}_{0}}e_{0}^{2} \right)}{{{k}_{B}}T} \right] . \label{eq:phi_min_estimation}
\end{equation}
Equation \eqref{eq:phi_min_estimation} suggests that the existence of a dilute phase $\bar{\phi} \approx 0$ requires that $a (h_{\phi} - k_0 e_0^2) / {k_B T} \ll 0$. 
\item For the dense phase, i.e., $\bar{\phi }\to 1$, (correspondingly $\bar{e}\to 0$), we have 
\begin{equation}
g\left( \bar{\phi }\to 1 \right)\simeq -\frac{{{k}_{B}}T}{a}\ln \left( 1-\bar{\phi } \right)-{{h}_{\phi }}+\kappa c_{0,\max}^2 = 0 . 
\end{equation}
We thus have the estimation of the dense phase as 
\begin{equation}
\bar{\phi }\simeq 1-\exp \left[ -\frac{a\left( {{h}_{\phi }}-\kappa c_{0,\max }^{2} \right)}{{{k}_{B}}T} \right] . \label{eq:phi_max_estimation}
\end{equation}
Equation \eqref{eq:phi_max_estimation} suggests that the existence of a dense phase $\bar{\phi} \approx 1$ requires that $h_{\phi} \gg \kappa c_{0,\max}^2 + {k_B T}/a$. 
\end{itemize}

\section{Approximation of the critical points} \label{sec:Analysis_h_phi_3}

\paragraph*{Triple point} We find that the \textit{triple point} is well approximated by the endpoint of the three homogeneous states region, see Fig. \ref{fig_NumberOfHomogeneousStates}.
Here we provide an analytical expression for such endpoint, denoted ($h_{\phi}^{\rm cr,1}$, $c_{0,\max}^{\rm cr,1}$). Letting $\min_{\phi \in (0,1)} \left\{ \frac{\mathrm{d}g}{\mathrm{d}\phi} \right\} = 0$ in Eq. \eqref{eq:min_dg_dphi}, we obtain 
\begin{equation}
c_{0,\max}^{\rm cr,1} = \sqrt{\frac{k_0 e_0^2}{\kappa} - \frac{4 k_B T}{a\kappa}} . 
\end{equation}
In addition, under the condition $g(\bar{\phi} = 1/2) = 0$, we find that:
\begin{equation}
h_{\phi}^{\rm cr,1} = \frac{1}{2} \left[ k_0 e_0^2 + \kappa ( c_{0,\max}^{\rm cr,1} )^2 \right] = k_0 e_0^2 - 2\frac{k_B T}{a} . 
\end{equation}

\paragraph*{Critical point} The critical point ($h_{\phi}^{\rm cr,2}$, $c_{0,\max}^{\rm cr,2}$) is defined by the conditions $\bar{\phi} = 1/2$ and $\beta_{\min} = 0$. With $\bar{\phi} = 1/2$, the coefficients $\beta_i$ ($i = 0,1,2,3$) read
\begin{equation}
\begin{split}
{{\beta }_{0}} = & \ 4\frac{{{k}_{0}e_0^2}{{k}_{B}}T}{a} + {{k}_{0}e_0^2}\kappa c_{0,\max }^{2} - k_{0}^{2}e_{0}^{4} , \\ 
{{\beta }_{1}} = & \ 4\frac{\sigma e_0^2 {{k}_{B}}T}{a} + {{k}_{0}e_0^2}{{D}_{\phi }} + \sigma e_0^2 \kappa c_{0,\max }^{2} \\
& \ - 2{{k}_{0}}{{e}_{0}^3} \kappa {{c}_{0,\max }} , \\ 
{{\beta }_{2}} = & \ 4\frac{\kappa e_0^2 {{k}_{B}}T}{a} + \sigma e_0^2 {{D}_{\phi }} , \\ 
{{\beta }_{3}} = & \ \kappa e_0^2 {{D}_{\phi }} , 
\end{split}
\end{equation}
The $\beta_{\min} = 0$ condition reads 
\begin{equation}
\beta_0 + \beta_1 q_s^2 + \beta_2 q_s^4 + \beta_3 q_s^6 = 0 , \label{eq:beta_min_zero}
\end{equation}
where $q_s$ is given by Eq. \eqref{eq:qs}. Solving Eq. \eqref{eq:beta_min_zero} gives the critical point ($h_{\phi}^{\rm cr,2}$, $c_{0,\max}^{\rm cr,2}$). 

In the limit of $D_{\phi} = 0$, the $\beta_{\min} = 0$ condition simplifies to 
\begin{align}
  & {{\sigma }^{2}}{{\kappa }^{2}}c_{0,\max }^{4}-4\sigma {{k}_{0}}{{e}_{0}}{{\kappa }^{2}}c_{0,\max }^{3} \notag \\ 
 & +\left( 4k_{0}^{2}e_{0}^{2}\kappa +8{{\sigma }^{2}}\frac{{{k}_{B}}T}{a}-16{{k}_{0}}\kappa\frac{{{k}_{B}}T}{a}  \right)\kappa c_{0,\max }^{2} \notag \\ 
 & -16\sigma {{k}_{0}}{{e}_{0}}\kappa\frac{{{k}_{B}}T}{a} {{c}_{0,\max }} \notag \\ 
 & +16k_{0}^{2}e_{0}^{2}\kappa \frac{{{k}_{B}}T}{a}+\left( 16{{\sigma }^{2}}-64{{k}_{0}}\kappa  \right){{\left( \frac{{{k}_{B}}T}{a} \right)}^{2}}=0. \label{eq:c0max_cr_2_sigmaZero}
\end{align}
Solving for $c_{0,\max}$, the  root of the above equation corresponds to the critical value $c_{0,\max}^{\rm cr,2}$. 
In the limit $\sigma \to 0$, Eq. \eqref{eq:c0max_cr_2_sigmaZero} further simplifies to 
\begin{equation}
{{\left( \sigma {{c}_{0,\max }}-2{{k}_{0}}{{e}_{0}} \right)}^{2}} \approx 16\frac{{{k}_{0}}{{k}_{B}}T}{a} , 
\end{equation}
which leads to the critical value, 
\begin{equation}
c_{0,\max }^{\text{cr},2}\approx \frac{2{{k}_{0}}{{e}_{0}} \pm 4\sqrt{\dfrac{{{k}_{0}}{{k}_{B}}T}{a}}}{\sigma }\ \ \ ,\ \ \ \left( \sigma \to 0 \right) . 
\end{equation}
Based on the upper bound estimation of $c_{0,\max}$ for the cluster formation region, see Eq. \eqref{eq:c0max_cr2_upper_bound}, $c_{0,\max}^{\rm cr,2}$ should satisfy $c_{0,\max}^{\rm cr,2} < 2 k_0 e_0 / \sigma$ ($\sigma \to 0$). We thus exclude the larger root ($+$ sign) and obtain, 
\begin{equation}
c_{0,\max }^{\text{cr},2}\approx \frac{2{{k}_{0}}{{e}_{0}} - 4\sqrt{\dfrac{{{k}_{0}}{{k}_{B}}T}{a}}}{\sigma }\ \ \ ,\ \ \ \left( \sigma \to 0 \right) . 
\end{equation}
The critical value $h_{\phi}^{\rm cr,2}$ is related to $c_{0,\max}^{\rm cr,2}$: 
\begin{align}
h_{\phi }^{\text{cr},2} & = \frac{1}{2}\left[ {{k}_{0}}e_{0}^{2}+\kappa {{\left( c_{0,\max }^{\text{cr},2} \right)}^{2}} \right] \notag \\ 
& \approx \frac{1}{2}\left[ 1+\frac{4\kappa {{k}_{0}}}{{{\sigma }^{2}}}{{\left( 1-2\sqrt{\frac{{{k}_{B}}T}{a{{k}_{0}}e_{0}^{2}}} \right)}^{2}} \right]{{k}_{0}}e_{0}^{2} . 
\end{align}

Further, the condition $c_{0,\max}^{\rm cr,1} = c_{0,\max}^{\rm cr,2}$ yields the following critical rest-length height
\begin{equation}
e_0^{\rm cr} = 2\left( \frac{4\kappa {{k}_{0}}+{{\sigma }^{2}}}{4\kappa {{k}_{0}}-{{\sigma }^{2}}} \right)\sqrt{\frac{{{k}_{B}}T}{a{{k}_{0}}}}\approx 2\sqrt{\frac{{{k}_{B}}T}{a{{k}_{0}}}}\ \ \ \left( \sigma \to 0 \right) , 
\end{equation}
and critical membrane--substrate binding stiffness 
\begin{equation}
k_0^{\rm cr} \approx 4\frac{k_B T}{a e_0^2}\ \ \ \left( \sigma \to 0 \right) , 
\end{equation}
below which the homogeneous steady states are always stable. In agreement with these predictions, we find that no stable clusters form for $e_0 < e_0^{\rm cr}$ or $k_0 < k_0^{\rm cr}$ in our numerical simulations, see Fig. \ref{fig_e0_regulation}.

\section{Link to the Swift--Hohenberg theory} \label{sec:Link_to_SH_theory}

To better illustrate the connection of our theory to the Swift--Hohenberg one, we here focus on a simple case where we impose that $e = e_0 (1-\phi)$; such an ansatz is motivated by the form of the homogeneous state Eq. \eqref{eq_HomogeneousSteadyStateEquation_1}. 
In such a simplified case, the free energy density reduces to
\begin{align}
f = & \ \frac{{{k}_{B}}T}{a}\left[ \phi \ln \phi +\left( 1-\phi  \right)\ln \left( 1-\phi  \right) \right] \notag \\ 
& -\frac{1}{2}{{k}_{0}}e_{0}^{2}{{\left( 1-\phi  \right)}^{2}}+\frac{1}{2}\kappa {{\left( {{e}_{0}}{{\nabla }^{2}}\phi +{{c}_{0,\max }}\phi  \right)}^{2}} \notag \\ 
& -{{h}_{\phi }}\phi +\frac{1}{2}\left( {{D}_{\phi }}+\sigma e_{0}^{2} \right){{\left( \nabla \phi  \right)}^{2}} . 
\end{align}

The evolution equation of $\phi$, i.e., $\partial\phi / \partial t = - \delta F/ \delta\phi$, can be expressed as
\begin{align}
\frac{\partial \phi }{\partial t} = & \left( {{k}_{0}}e_{0}^{2}-\kappa c_{0,\max }^{2} \right)\phi-\kappa e_{0}^{2}{{\nabla }^{2}}{{\nabla }^{2}}\phi \notag \\
& +\left( {{D}_{\phi }}+\sigma e_{0}^{2}-2\kappa {{e}_{0}}{{c}_{0,\max }} \right){{\nabla }^{2}}\phi \notag \\
& +{{h}_{\phi }}-{{k}_{0}}e_{0}^{2}-\frac{{{k}_{B}}T}{a}\ln \left( \frac{\phi }{1-\phi } \right) . \label{eq:phi_simple_model}
\end{align}
Letting $\delta\phi = \phi - \bar{\phi}$ with $\bar{\phi}$ being the homogeneous state given by Eq. \eqref{eq_HomogeneousSteadyStateEquation_1} and expanding Eq. \eqref{eq:phi_simple_model} to third-order terms in $\delta \phi$, we find that: 
\begin{align}
\frac{\partial \delta\phi }{\partial t} = & \left[ {{k}_{0}}e_{0}^{2}-\kappa c_{0,\max }^{2} - \frac{k_B T}{a \bar{\phi} (1-\bar{\phi})} \right] \delta\phi  \notag \\
& -\left( 2\kappa {{e}_{0}}{{c}_{0,\max }} - \sigma e_{0}^{2} - {{D}_{\phi }} \right){{\nabla }^{2}}\delta\phi \notag \\
& - \kappa e_{0}^{2}{{\nabla }^{2}}{{\nabla }^{2}}\delta\phi + n (\delta\phi) ,  \label{eq:phi_simple_model_2}
\end{align}
where
\begin{align}
n(\delta\phi) = \ & \frac{{{k}_{B}}T}{2 a } \frac{(1-2 \bar{\phi})}{\bar{\phi}^2(1-\bar{\phi})^2 } (\delta \phi)^2 \notag \\
& - \frac{{{k}_{B}}T}{3 a } \frac{1 - 3 \bar{\phi} (1-\bar{\phi})}{\bar{\phi}^3 (1-\bar{\phi})^3} (\delta \phi)^3.
\end{align}
When the spontaneous curvature $c_{0,\max}$ is large, i.e., 
\begin{equation}
c_{0,\max} > \frac{D_{\phi} + \sigma e_0^2}{2 \kappa e_0} \triangleq c_{0,\max}^{\rm th}, 
\end{equation}
Eq. \eqref{eq:phi_simple_model_2} can be recast as a Swift--Hohenberg equation, see Eq. \eqref{eq:Swift-Hohenberg}. 


\section{Simulation scheme} \label{sec:Simulation_scheme}

We use the spectral method to solve the controlling equations \eqref{eq_dedt} and \eqref{eq_dphidt}. The time integration is performed using a backward Euler scheme; the spatial derivatives are carried out using a second-order central difference method. 
Simulations were performed on a $256 \times 256$ two-dimensional lattice using periodic boundary conditions. 

To converge to the energy minimum, we consider the following gradient-descent dynamics, $\dot{\phi}=-\delta F/\delta \phi$, and $\dot{e} = -\delta F/\delta e +\eta(\bm{x}, t)$, i.e., Eqs. \eqref{eq_dedt} and \eqref{eq_dphidt}; the white noise is discretized as, 
\begin{equation}
\eta \left( \bm{x},t \right)=\frac{\Lambda }{\sqrt{{{\left( \Delta x \right)}^{2}}\Delta t}}\vartheta , 
\end{equation}
where $\vartheta$ is the standard normal distribution, $\Delta x$ and $\Delta t$ are the space step and the time step, respectively. 
We set $\Delta x = 1/16$ and $\Delta t = 10^{-6}$ (non-dimensional values) in our simulations. 
To approach the global energy minimum state, we performed annealing simulations where we decrease the noise intensity $\Lambda$ gradually from $\Lambda = 10$ to $\Lambda = 0$. We decrease $\Lambda$ quasi-statically, according to the iterative process: (1) relaxing the system to reach a steady state with a noise level $\Lambda^{(0)} = 10$; (2) reduce the noise intensity to $\Lambda^{(i+1)} = \Lambda^{(i)} + \Delta \Lambda$ with $\Delta \Lambda = -0.1$. We repeat the above two steps until $\Lambda = 0$. 

The initial conditions for the fields $e$ and $\phi$ correspond to that of a homogeneous steady state with small perturbations, $e(\bm{x} , t = 0) = \bar{e} + \varepsilon_{e} \vartheta_{e} (\bm{x})$ and $\phi(\bm{x} , t = 0) = \bar{\phi} + \varepsilon_{\phi} \vartheta_{\phi} (\bm{x})$, where $\varepsilon_{e} = 10^{-3}$ and $\varepsilon_{\phi} = 10^{-3}$; $\vartheta_{e} (\bm{x})$ and $\vartheta_{\phi} (\bm{x})$ are random valuables satisfying the normal distribution. We checked that the final steady state does not depend on the particular choice of the initial state; this suggests that the global free energy minimum is reached.


%


\begin{thebibliography}{38}%
\makeatletter
\providecommand \@ifxundefined [1]{%
 \@ifx{#1\undefined}
}%
\providecommand \@ifnum [1]{%
 \ifnum #1\expandafter \@firstoftwo
 \else \expandafter \@secondoftwo
 \fi
}%
\providecommand \@ifx [1]{%
 \ifx #1\expandafter \@firstoftwo
 \else \expandafter \@secondoftwo
 \fi
}%
\providecommand \natexlab [1]{#1}%
\providecommand \enquote  [1]{``#1''}%
\providecommand \bibnamefont  [1]{#1}%
\providecommand \bibfnamefont [1]{#1}%
\providecommand \citenamefont [1]{#1}%
\providecommand \href@noop [0]{\@secondoftwo}%
\providecommand \href [0]{\begingroup \@sanitize@url \@href}%
\providecommand \@href[1]{\@@startlink{#1}\@@href}%
\providecommand \@@href[1]{\endgroup#1\@@endlink}%
\providecommand \@sanitize@url [0]{\catcode `\\12\catcode `\$12\catcode
  `\&12\catcode `\#12\catcode `\^12\catcode `\_12\catcode `\%12\relax}%
\providecommand \@@startlink[1]{}%
\providecommand \@@endlink[0]{}%
\providecommand \url  [0]{\begingroup\@sanitize@url \@url }%
\providecommand \@url [1]{\endgroup\@href {#1}{\urlprefix }}%
\providecommand \urlprefix  [0]{URL }%
\providecommand \Eprint [0]{\href }%
\providecommand \doibase [0]{https://doi.org/}%
\providecommand \selectlanguage [0]{\@gobble}%
\providecommand \bibinfo  [0]{\@secondoftwo}%
\providecommand \bibfield  [0]{\@secondoftwo}%
\providecommand \translation [1]{[#1]}%
\providecommand \BibitemOpen [0]{}%
\providecommand \bibitemStop [0]{}%
\providecommand \bibitemNoStop [0]{.\EOS\space}%
\providecommand \EOS [0]{\spacefactor3000\relax}%
\providecommand \BibitemShut  [1]{\csname bibitem#1\endcsname}%
\let\auto@bib@innerbib\@empty
\bibitem [{\citenamefont {Gumbiner}(1996)}]{Gumbiner1996}%
  \BibitemOpen
  \bibfield  {author} {\bibinfo {author} {\bibfnamefont {B.~M.}\ \bibnamefont
  {Gumbiner}},\ }\href {https://doi.org/10.1016/S0092-8674(00)81279-9}
  {\bibfield  {journal} {\bibinfo  {journal} {Cell}\ }\textbf {\bibinfo
  {volume} {84}},\ \bibinfo {pages} {345} (\bibinfo {year} {1996})}\BibitemShut
  {NoStop}%
\bibitem [{\citenamefont {Ladoux}\ and\ \citenamefont
  {Nicolas}(2012)}]{Ladoux2012}%
  \BibitemOpen
  \bibfield  {author} {\bibinfo {author} {\bibfnamefont {B.}~\bibnamefont
  {Ladoux}}\ and\ \bibinfo {author} {\bibfnamefont {A.}~\bibnamefont
  {Nicolas}},\ }\href {https://doi.org/10.1088/0034-4885/75/11/116601}
  {\bibfield  {journal} {\bibinfo  {journal} {Reports on Progress in Physics}\
  }\textbf {\bibinfo {volume} {75}},\ \bibinfo {pages} {116601} (\bibinfo
  {year} {2012})}\BibitemShut {NoStop}%
\bibitem [{\citenamefont {Schwarz}\ and\ \citenamefont
  {Safran}(2013)}]{Schwarz2013}%
  \BibitemOpen
  \bibfield  {author} {\bibinfo {author} {\bibfnamefont {U.~S.}\ \bibnamefont
  {Schwarz}}\ and\ \bibinfo {author} {\bibfnamefont {S.~A.}\ \bibnamefont
  {Safran}},\ }\href {https://doi.org/10.1103/RevModPhys.85.1327} {\bibfield
  {journal} {\bibinfo  {journal} {Reviews of Modern Physics}\ }\textbf
  {\bibinfo {volume} {85}},\ \bibinfo {pages} {1327} (\bibinfo {year}
  {2013})}\BibitemShut {NoStop}%
\bibitem [{\citenamefont {Sun}\ \emph {et~al.}(2019)\citenamefont {Sun},
  \citenamefont {Costell},\ and\ \citenamefont {F{\"a}ssler}}]{Sun2019}%
  \BibitemOpen
  \bibfield  {author} {\bibinfo {author} {\bibfnamefont {Z.}~\bibnamefont
  {Sun}}, \bibinfo {author} {\bibfnamefont {M.}~\bibnamefont {Costell}},\ and\
  \bibinfo {author} {\bibfnamefont {R.}~\bibnamefont {F{\"a}ssler}},\ }\href
  {https://doi.org/10.1038/s41556-018-0234-9} {\bibfield  {journal} {\bibinfo
  {journal} {Nature Cell Biology}\ }\textbf {\bibinfo {volume} {21}},\ \bibinfo
  {pages} {25} (\bibinfo {year} {2019})}\BibitemShut {NoStop}%
\bibitem [{\citenamefont {Janiszewska}\ \emph {et~al.}(2020)\citenamefont
  {Janiszewska}, \citenamefont {Primi},\ and\ \citenamefont
  {Izard}}]{Janiszewska2020}%
  \BibitemOpen
  \bibfield  {author} {\bibinfo {author} {\bibfnamefont {M.}~\bibnamefont
  {Janiszewska}}, \bibinfo {author} {\bibfnamefont {M.~C.}\ \bibnamefont
  {Primi}},\ and\ \bibinfo {author} {\bibfnamefont {T.}~\bibnamefont {Izard}},\
  }\href {https://doi.org/10.1074/jbc.REV119.007759} {\bibfield  {journal}
  {\bibinfo  {journal} {Journal of Biological Chemistry}\ }\textbf {\bibinfo
  {volume} {295}},\ \bibinfo {pages} {2495} (\bibinfo {year}
  {2020})}\BibitemShut {NoStop}%
\bibitem [{\citenamefont {Changede}\ \emph {et~al.}(2015)\citenamefont
  {Changede}, \citenamefont {Xu}, \citenamefont {Margadant},\ and\
  \citenamefont {Sheetz}}]{Changede2015}%
  \BibitemOpen
  \bibfield  {author} {\bibinfo {author} {\bibfnamefont {R.}~\bibnamefont
  {Changede}}, \bibinfo {author} {\bibfnamefont {X.}~\bibnamefont {Xu}},
  \bibinfo {author} {\bibfnamefont {F.}~\bibnamefont {Margadant}},\ and\
  \bibinfo {author} {\bibfnamefont {M.}~\bibnamefont {Sheetz}},\ }\href
  {https://doi.org/10.1016/j.devcel.2015.11.001} {\bibfield  {journal}
  {\bibinfo  {journal} {Developmental Cell}\ }\textbf {\bibinfo {volume}
  {35}},\ \bibinfo {pages} {614} (\bibinfo {year} {2015})}\BibitemShut
  {NoStop}%
\bibitem [{\citenamefont {Yu}\ \emph {et~al.}(2015)\citenamefont {Yu},
  \citenamefont {Rafiq}, \citenamefont {Cao}, \citenamefont {Zhou},
  \citenamefont {Krishnasamy}, \citenamefont {Biswas}, \citenamefont {Ravasio},
  \citenamefont {Chen}, \citenamefont {Wang}, \citenamefont {Kawauchi} \emph
  {et~al.}}]{Yu2015}%
  \BibitemOpen
  \bibfield  {author} {\bibinfo {author} {\bibfnamefont {C.-h.}\ \bibnamefont
  {Yu}}, \bibinfo {author} {\bibfnamefont {N.~B.~M.}\ \bibnamefont {Rafiq}},
  \bibinfo {author} {\bibfnamefont {F.}~\bibnamefont {Cao}}, \bibinfo {author}
  {\bibfnamefont {Y.}~\bibnamefont {Zhou}}, \bibinfo {author} {\bibfnamefont
  {A.}~\bibnamefont {Krishnasamy}}, \bibinfo {author} {\bibfnamefont {K.~H.}\
  \bibnamefont {Biswas}}, \bibinfo {author} {\bibfnamefont {A.}~\bibnamefont
  {Ravasio}}, \bibinfo {author} {\bibfnamefont {Z.}~\bibnamefont {Chen}},
  \bibinfo {author} {\bibfnamefont {Y.-H.}\ \bibnamefont {Wang}}, \bibinfo
  {author} {\bibfnamefont {K.}~\bibnamefont {Kawauchi}}, \emph {et~al.},\
  }\href {https://doi.org/10.1038/ncomms9672} {\bibfield  {journal} {\bibinfo
  {journal} {Nature Communications}\ }\textbf {\bibinfo {volume} {6}},\
  \bibinfo {pages} {8672} (\bibinfo {year} {2015})}\BibitemShut {NoStop}%
\bibitem [{\citenamefont {Changede}\ \emph {et~al.}(2019)\citenamefont
  {Changede}, \citenamefont {Cai}, \citenamefont {Wind},\ and\ \citenamefont
  {Sheetz}}]{Changede2019}%
  \BibitemOpen
  \bibfield  {author} {\bibinfo {author} {\bibfnamefont {R.}~\bibnamefont
  {Changede}}, \bibinfo {author} {\bibfnamefont {H.}~\bibnamefont {Cai}},
  \bibinfo {author} {\bibfnamefont {S.~J.}\ \bibnamefont {Wind}},\ and\
  \bibinfo {author} {\bibfnamefont {M.~P.}\ \bibnamefont {Sheetz}},\ }\href
  {https://doi.org/10.1038/s41563-019-0460-y} {\bibfield  {journal} {\bibinfo
  {journal} {Nature Materials}\ }\textbf {\bibinfo {volume} {18}},\ \bibinfo
  {pages} {1366} (\bibinfo {year} {2019})}\BibitemShut {NoStop}%
\bibitem [{\citenamefont {Lin}\ \emph {et~al.}(2023)\citenamefont {Lin},
  \citenamefont {Changede}, \citenamefont {Sheetz}, \citenamefont {Prost},\
  and\ \citenamefont {Rupprecht}}]{Lin2023}%
  \BibitemOpen
  \bibfield  {author} {\bibinfo {author} {\bibfnamefont {S.~Z.}\ \bibnamefont
  {Lin}}, \bibinfo {author} {\bibfnamefont {R.}~\bibnamefont {Changede}},
  \bibinfo {author} {\bibfnamefont {M.~P.}\ \bibnamefont {Sheetz}}, \bibinfo
  {author} {\bibfnamefont {J.}~\bibnamefont {Prost}},\ and\ \bibinfo {author}
  {\bibfnamefont {J.-F.}\ \bibnamefont {Rupprecht}},\ }\href@noop {} {\bibinfo
  {title} {Tilt-induced clustering of cell adhesion proteins}} (\bibinfo {year}
  {2023}),\ \Eprint {https://arxiv.org/abs/2307.03670} {arXiv:2307.03670
  [cond-mat.soft]} \BibitemShut {NoStop}%
\bibitem [{\citenamefont {Marcerou}\ \emph {et~al.}(1984)\citenamefont
  {Marcerou}, \citenamefont {Prost},\ and\ \citenamefont
  {Gruler}}]{Marcerou1984}%
  \BibitemOpen
  \bibfield  {author} {\bibinfo {author} {\bibfnamefont {J.~P.}\ \bibnamefont
  {Marcerou}}, \bibinfo {author} {\bibfnamefont {J.}~\bibnamefont {Prost}},\
  and\ \bibinfo {author} {\bibfnamefont {H.}~\bibnamefont {Gruler}},\ }\href
  {https://doi.org/10.1007/BF02452212} {\bibfield  {journal} {\bibinfo
  {journal} {Il Nuovo Cimento D}\ }\textbf {\bibinfo {volume} {3}},\ \bibinfo
  {pages} {204} (\bibinfo {year} {1984})}\BibitemShut {NoStop}%
\bibitem [{\citenamefont {Safran}(1999)}]{doi:10.1080/000187399243428}%
  \BibitemOpen
  \bibfield  {author} {\bibinfo {author} {\bibfnamefont {S.~A.}\ \bibnamefont
  {Safran}},\ }\href {https://doi.org/10.1080/000187399243428} {\bibfield
  {journal} {\bibinfo  {journal} {Advances in Physics}\ }\textbf {\bibinfo
  {volume} {48}},\ \bibinfo {pages} {395} (\bibinfo {year} {1999})}\BibitemShut
  {NoStop}%
\bibitem [{\citenamefont {Sheetz}\ and\ \citenamefont
  {Singer}(1974)}]{Sheetz1974}%
  \BibitemOpen
  \bibfield  {author} {\bibinfo {author} {\bibfnamefont {M.~P.}\ \bibnamefont
  {Sheetz}}\ and\ \bibinfo {author} {\bibfnamefont {S.}~\bibnamefont
  {Singer}},\ }\href {https://doi.org/10.1073/pnas.71.11.4457} {\bibfield
  {journal} {\bibinfo  {journal} {Proceedings of the National Academy of
  Sciences of the United States of America}\ }\textbf {\bibinfo {volume}
  {71}},\ \bibinfo {pages} {4457} (\bibinfo {year} {1974})}\BibitemShut
  {NoStop}%
\bibitem [{\citenamefont {Stachowiak}\ \emph {et~al.}(2012)\citenamefont
  {Stachowiak}, \citenamefont {Schmid}, \citenamefont {Ryan}, \citenamefont
  {Ann}, \citenamefont {Sasaki}, \citenamefont {Sherman}, \citenamefont
  {Geissler}, \citenamefont {Fletcher},\ and\ \citenamefont
  {Hayden}}]{Stachowiak2012}%
  \BibitemOpen
  \bibfield  {author} {\bibinfo {author} {\bibfnamefont {J.~C.}\ \bibnamefont
  {Stachowiak}}, \bibinfo {author} {\bibfnamefont {E.~M.}\ \bibnamefont
  {Schmid}}, \bibinfo {author} {\bibfnamefont {C.~J.}\ \bibnamefont {Ryan}},
  \bibinfo {author} {\bibfnamefont {H.~S.}\ \bibnamefont {Ann}}, \bibinfo
  {author} {\bibfnamefont {D.~Y.}\ \bibnamefont {Sasaki}}, \bibinfo {author}
  {\bibfnamefont {M.~B.}\ \bibnamefont {Sherman}}, \bibinfo {author}
  {\bibfnamefont {P.~L.}\ \bibnamefont {Geissler}}, \bibinfo {author}
  {\bibfnamefont {D.~A.}\ \bibnamefont {Fletcher}},\ and\ \bibinfo {author}
  {\bibfnamefont {C.~C.}\ \bibnamefont {Hayden}},\ }\href
  {https://doi.org/10.1038/ncb2561} {\bibfield  {journal} {\bibinfo  {journal}
  {Nature Cell Biology}\ }\textbf {\bibinfo {volume} {14}},\ \bibinfo {pages}
  {944} (\bibinfo {year} {2012})}\BibitemShut {NoStop}%
\bibitem [{\citenamefont {McMahon}\ and\ \citenamefont
  {Boucrot}(2015)}]{McMahon2015}%
  \BibitemOpen
  \bibfield  {author} {\bibinfo {author} {\bibfnamefont {H.~T.}\ \bibnamefont
  {McMahon}}\ and\ \bibinfo {author} {\bibfnamefont {E.}~\bibnamefont
  {Boucrot}},\ }\href {https://doi.org/10.1242/jcs.114454} {\bibfield
  {journal} {\bibinfo  {journal} {Journal of Cell Science}\ }\textbf {\bibinfo
  {volume} {128}},\ \bibinfo {pages} {1065} (\bibinfo {year}
  {2015})}\BibitemShut {NoStop}%
\bibitem [{\citenamefont {Ramakrishnan}\ \emph {et~al.}(2013)\citenamefont
  {Ramakrishnan}, \citenamefont {Kumar},\ and\ \citenamefont
  {Ipsen}}]{Ramakrishnan2013}%
  \BibitemOpen
  \bibfield  {author} {\bibinfo {author} {\bibfnamefont {N.}~\bibnamefont
  {Ramakrishnan}}, \bibinfo {author} {\bibfnamefont {P.~S.}\ \bibnamefont
  {Kumar}},\ and\ \bibinfo {author} {\bibfnamefont {J.~H.}\ \bibnamefont
  {Ipsen}},\ }\href {https://doi.org/10.1016/j.bpj.2012.12.045} {\bibfield
  {journal} {\bibinfo  {journal} {Biophysical Journal}\ }\textbf {\bibinfo
  {volume} {104}},\ \bibinfo {pages} {1018} (\bibinfo {year}
  {2013})}\BibitemShut {NoStop}%
\bibitem [{\citenamefont {Zakany}\ \emph {et~al.}(2020)\citenamefont {Zakany},
  \citenamefont {Kovacs}, \citenamefont {Panyi},\ and\ \citenamefont
  {Varga}}]{Zakany2020}%
  \BibitemOpen
  \bibfield  {author} {\bibinfo {author} {\bibfnamefont {F.}~\bibnamefont
  {Zakany}}, \bibinfo {author} {\bibfnamefont {T.}~\bibnamefont {Kovacs}},
  \bibinfo {author} {\bibfnamefont {G.}~\bibnamefont {Panyi}},\ and\ \bibinfo
  {author} {\bibfnamefont {Z.}~\bibnamefont {Varga}},\ }\href
  {https://doi.org/10.1016/j.bbalip.2020.158706} {\bibfield  {journal}
  {\bibinfo  {journal} {Biochimica et Biophysica Acta (BBA)-Molecular and Cell
  Biology of Lipids}\ }\textbf {\bibinfo {volume} {1865}},\ \bibinfo {pages}
  {158706} (\bibinfo {year} {2020})}\BibitemShut {NoStop}%
\bibitem [{\citenamefont {Zamir}\ \emph {et~al.}(2000)\citenamefont {Zamir},
  \citenamefont {Katz}, \citenamefont {Posen}, \citenamefont {Erez},
  \citenamefont {Yamada}, \citenamefont {Katz}, \citenamefont {Lin},
  \citenamefont {Lin}, \citenamefont {Bershadsky}, \citenamefont {Kam},\ and\
  \citenamefont {Geiger}}]{Zamir2000}%
  \BibitemOpen
  \bibfield  {author} {\bibinfo {author} {\bibfnamefont {E.}~\bibnamefont
  {Zamir}}, \bibinfo {author} {\bibfnamefont {M.}~\bibnamefont {Katz}},
  \bibinfo {author} {\bibfnamefont {Y.}~\bibnamefont {Posen}}, \bibinfo
  {author} {\bibfnamefont {N.}~\bibnamefont {Erez}}, \bibinfo {author}
  {\bibfnamefont {K.~M.}\ \bibnamefont {Yamada}}, \bibinfo {author}
  {\bibfnamefont {B.-Z.}\ \bibnamefont {Katz}}, \bibinfo {author}
  {\bibfnamefont {S.}~\bibnamefont {Lin}}, \bibinfo {author} {\bibfnamefont
  {D.~C.}\ \bibnamefont {Lin}}, \bibinfo {author} {\bibfnamefont
  {A.}~\bibnamefont {Bershadsky}}, \bibinfo {author} {\bibfnamefont
  {Z.}~\bibnamefont {Kam}},\ and\ \bibinfo {author} {\bibfnamefont
  {B.}~\bibnamefont {Geiger}},\ }\href {https://doi.org/10.1038/35008607}
  {\bibfield  {journal} {\bibinfo  {journal} {Nature Cell Biology}\ }\textbf
  {\bibinfo {volume} {2}},\ \bibinfo {pages} {191} (\bibinfo {year}
  {2000})}\BibitemShut {NoStop}%
\bibitem [{\citenamefont {Bihr}\ \emph {et~al.}(2012)\citenamefont {Bihr},
  \citenamefont {Seifert},\ and\ \citenamefont {Smith}}]{Bihr2012}%
  \BibitemOpen
  \bibfield  {author} {\bibinfo {author} {\bibfnamefont {T.}~\bibnamefont
  {Bihr}}, \bibinfo {author} {\bibfnamefont {U.}~\bibnamefont {Seifert}},\ and\
  \bibinfo {author} {\bibfnamefont {A.-S.}\ \bibnamefont {Smith}},\ }\href
  {https://doi.org/10.1103/PhysRevLett.109.258101} {\bibfield  {journal}
  {\bibinfo  {journal} {Physical Review Letters}\ }\textbf {\bibinfo {volume}
  {109}},\ \bibinfo {pages} {258101} (\bibinfo {year} {2012})}\BibitemShut
  {NoStop}%
\bibitem [{\citenamefont {Bihr}\ \emph {et~al.}(2015)\citenamefont {Bihr},
  \citenamefont {Seifert},\ and\ \citenamefont {Smith}}]{Bihr2015}%
  \BibitemOpen
  \bibfield  {author} {\bibinfo {author} {\bibfnamefont {T.}~\bibnamefont
  {Bihr}}, \bibinfo {author} {\bibfnamefont {U.}~\bibnamefont {Seifert}},\ and\
  \bibinfo {author} {\bibfnamefont {A.-S.}\ \bibnamefont {Smith}},\ }\href
  {https://doi.org/10.1088/1367-2630/17/8/083016} {\bibfield  {journal}
  {\bibinfo  {journal} {New Journal of Physics}\ }\textbf {\bibinfo {volume}
  {17}},\ \bibinfo {pages} {083016} (\bibinfo {year} {2015})}\BibitemShut
  {NoStop}%
\bibitem [{\citenamefont {Li}\ \emph {et~al.}(2019)\citenamefont {Li},
  \citenamefont {Hu}, \citenamefont {Li},\ and\ \citenamefont {Song}}]{Li2019}%
  \BibitemOpen
  \bibfield  {author} {\bibinfo {author} {\bibfnamefont {L.}~\bibnamefont
  {Li}}, \bibinfo {author} {\bibfnamefont {J.}~\bibnamefont {Hu}}, \bibinfo
  {author} {\bibfnamefont {L.}~\bibnamefont {Li}},\ and\ \bibinfo {author}
  {\bibfnamefont {F.}~\bibnamefont {Song}},\ }\href {DOI
  https://doi.org/10.1039/C8SM02504E} {\bibfield  {journal} {\bibinfo
  {journal} {Soft Matter}\ }\textbf {\bibinfo {volume} {15}},\ \bibinfo {pages}
  {3507} (\bibinfo {year} {2019})}\BibitemShut {NoStop}%
\bibitem [{\citenamefont {Huggins}(1941)}]{Huggins1941}%
  \BibitemOpen
  \bibfield  {author} {\bibinfo {author} {\bibfnamefont {M.~L.}\ \bibnamefont
  {Huggins}},\ }\href {https://doi.org/10.1063/1.1750930} {\bibfield  {journal}
  {\bibinfo  {journal} {The Journal of Chemical Physics}\ }\textbf {\bibinfo
  {volume} {9}},\ \bibinfo {pages} {440} (\bibinfo {year} {1941})}\BibitemShut
  {NoStop}%
\bibitem [{\citenamefont {Flory}(1941)}]{Flory1941}%
  \BibitemOpen
  \bibfield  {author} {\bibinfo {author} {\bibfnamefont {P.~J.}\ \bibnamefont
  {Flory}},\ }\href {https://doi.org/10.1063/1.1750971} {\bibfield  {journal}
  {\bibinfo  {journal} {The Journal of Chemical Physics}\ }\textbf {\bibinfo
  {volume} {9}},\ \bibinfo {pages} {660} (\bibinfo {year} {1941})}\BibitemShut
  {NoStop}%
\bibitem [{\citenamefont {Raote}\ \emph {et~al.}(2020)\citenamefont {Raote},
  \citenamefont {Chabanon}, \citenamefont {Walani}, \citenamefont {Arroyo},
  \citenamefont {Garcia-Parajo}, \citenamefont {Malhotra},\ and\ \citenamefont
  {Campelo}}]{Raote2020}%
  \BibitemOpen
  \bibfield  {author} {\bibinfo {author} {\bibfnamefont {I.}~\bibnamefont
  {Raote}}, \bibinfo {author} {\bibfnamefont {M.}~\bibnamefont {Chabanon}},
  \bibinfo {author} {\bibfnamefont {N.}~\bibnamefont {Walani}}, \bibinfo
  {author} {\bibfnamefont {M.}~\bibnamefont {Arroyo}}, \bibinfo {author}
  {\bibfnamefont {M.~F.}\ \bibnamefont {Garcia-Parajo}}, \bibinfo {author}
  {\bibfnamefont {V.}~\bibnamefont {Malhotra}},\ and\ \bibinfo {author}
  {\bibfnamefont {F.}~\bibnamefont {Campelo}},\ }\href
  {https://doi.org/10.7554/eLife.59426} {\bibfield  {journal} {\bibinfo
  {journal} {eLife}\ }\textbf {\bibinfo {volume} {9}},\ \bibinfo {pages}
  {e59426} (\bibinfo {year} {2020})}\BibitemShut {NoStop}%
\bibitem [{\citenamefont {Gov}(2018)}]{Gov2018}%
  \BibitemOpen
  \bibfield  {author} {\bibinfo {author} {\bibfnamefont {N.~S.}\ \bibnamefont
  {Gov}},\ }\href {https://doi.org/10.1098/rstb.2017.0115} {\bibfield
  {journal} {\bibinfo  {journal} {Philosophical Transactions of the Royal
  Society B: Biological Sciences}\ }\textbf {\bibinfo {volume} {373}},\
  \bibinfo {pages} {20170115} (\bibinfo {year} {2018})}\BibitemShut {NoStop}%
\bibitem [{\citenamefont {Smith}\ \emph {et~al.}(2008)\citenamefont {Smith},
  \citenamefont {Sengupta}, \citenamefont {Goennenwein}, \citenamefont
  {Seifert},\ and\ \citenamefont {Sackmann}}]{Smith2008}%
  \BibitemOpen
  \bibfield  {author} {\bibinfo {author} {\bibfnamefont {A.-S.}\ \bibnamefont
  {Smith}}, \bibinfo {author} {\bibfnamefont {K.}~\bibnamefont {Sengupta}},
  \bibinfo {author} {\bibfnamefont {S.}~\bibnamefont {Goennenwein}}, \bibinfo
  {author} {\bibfnamefont {U.}~\bibnamefont {Seifert}},\ and\ \bibinfo {author}
  {\bibfnamefont {E.}~\bibnamefont {Sackmann}},\ }\href
  {https://doi.org/10.1073/pnas.0801706105} {\bibfield  {journal} {\bibinfo
  {journal} {Proceedings of the National Academy of Sciences of the United
  States of America}\ }\textbf {\bibinfo {volume} {105}},\ \bibinfo {pages}
  {6906} (\bibinfo {year} {2008})}\BibitemShut {NoStop}%
\bibitem [{\citenamefont {Weikl}(2018)}]{Weikl2018}%
  \BibitemOpen
  \bibfield  {author} {\bibinfo {author} {\bibfnamefont {T.~R.}\ \bibnamefont
  {Weikl}},\ }\href {https://doi.org/10.1146/annurev-physchem-052516-050637}
  {\bibfield  {journal} {\bibinfo  {journal} {Annual Review of Physical
  Chemistry}\ }\textbf {\bibinfo {volume} {69}},\ \bibinfo {pages} {521}
  (\bibinfo {year} {2018})}\BibitemShut {NoStop}%
\bibitem [{\citenamefont {Steink{\"u}hler}\ \emph {et~al.}(2019)\citenamefont
  {Steink{\"u}hler}, \citenamefont {Sezgin}, \citenamefont
  {Urban{\v{c}}i{\v{c}}}, \citenamefont {Eggeling},\ and\ \citenamefont
  {Dimova}}]{Steinkuhler2019}%
  \BibitemOpen
  \bibfield  {author} {\bibinfo {author} {\bibfnamefont {J.}~\bibnamefont
  {Steink{\"u}hler}}, \bibinfo {author} {\bibfnamefont {E.}~\bibnamefont
  {Sezgin}}, \bibinfo {author} {\bibfnamefont {I.}~\bibnamefont
  {Urban{\v{c}}i{\v{c}}}}, \bibinfo {author} {\bibfnamefont {C.}~\bibnamefont
  {Eggeling}},\ and\ \bibinfo {author} {\bibfnamefont {R.}~\bibnamefont
  {Dimova}},\ }\href {https://doi.org/10.1038/s42003-019-0583-3} {\bibfield
  {journal} {\bibinfo  {journal} {Communications Biology}\ }\textbf {\bibinfo
  {volume} {2}},\ \bibinfo {pages} {337} (\bibinfo {year} {2019})}\BibitemShut
  {NoStop}%
\bibitem [{\citenamefont {Popescu}\ \emph {et~al.}(2006)\citenamefont
  {Popescu}, \citenamefont {Ikeda}, \citenamefont {Goda}, \citenamefont
  {Best-Popescu}, \citenamefont {Laposata}, \citenamefont {Manley},
  \citenamefont {Dasari}, \citenamefont {Badizadegan},\ and\ \citenamefont
  {Feld}}]{Popescu2006}%
  \BibitemOpen
  \bibfield  {author} {\bibinfo {author} {\bibfnamefont {G.}~\bibnamefont
  {Popescu}}, \bibinfo {author} {\bibfnamefont {T.}~\bibnamefont {Ikeda}},
  \bibinfo {author} {\bibfnamefont {K.}~\bibnamefont {Goda}}, \bibinfo {author}
  {\bibfnamefont {C.~A.}\ \bibnamefont {Best-Popescu}}, \bibinfo {author}
  {\bibfnamefont {M.}~\bibnamefont {Laposata}}, \bibinfo {author}
  {\bibfnamefont {S.}~\bibnamefont {Manley}}, \bibinfo {author} {\bibfnamefont
  {R.~R.}\ \bibnamefont {Dasari}}, \bibinfo {author} {\bibfnamefont
  {K.}~\bibnamefont {Badizadegan}},\ and\ \bibinfo {author} {\bibfnamefont
  {M.~S.}\ \bibnamefont {Feld}},\ }\href
  {https://doi.org/10.1103/PhysRevLett.97.218101} {\bibfield  {journal}
  {\bibinfo  {journal} {Physical Review Letters}\ }\textbf {\bibinfo {volume}
  {97}},\ \bibinfo {pages} {218101} (\bibinfo {year} {2006})}\BibitemShut
  {NoStop}%
\bibitem [{\citenamefont {Kozlov}\ and\ \citenamefont
  {Chernomordik}(2015)}]{Kozlov2015}%
  \BibitemOpen
  \bibfield  {author} {\bibinfo {author} {\bibfnamefont {M.~M.}\ \bibnamefont
  {Kozlov}}\ and\ \bibinfo {author} {\bibfnamefont {L.~V.}\ \bibnamefont
  {Chernomordik}},\ }\href {https://doi.org/10.1016/j.sbi.2015.07.010}
  {\bibfield  {journal} {\bibinfo  {journal} {Current Opinion in Structural
  Biology}\ }\textbf {\bibinfo {volume} {33}},\ \bibinfo {pages} {61} (\bibinfo
  {year} {2015})}\BibitemShut {NoStop}%
\bibitem [{\citenamefont {Xu}\ \emph {et~al.}(2016)\citenamefont {Xu},
  \citenamefont {Kim}, \citenamefont {Swift}, \citenamefont {Smith},
  \citenamefont {Volkmann},\ and\ \citenamefont {Hanein}}]{Xu2016}%
  \BibitemOpen
  \bibfield  {author} {\bibinfo {author} {\bibfnamefont {X.-P.}\ \bibnamefont
  {Xu}}, \bibinfo {author} {\bibfnamefont {E.}~\bibnamefont {Kim}}, \bibinfo
  {author} {\bibfnamefont {M.}~\bibnamefont {Swift}}, \bibinfo {author}
  {\bibfnamefont {J.}~\bibnamefont {Smith}}, \bibinfo {author} {\bibfnamefont
  {N.}~\bibnamefont {Volkmann}},\ and\ \bibinfo {author} {\bibfnamefont
  {D.}~\bibnamefont {Hanein}},\ }\href
  {https://doi.org/https://doi.org/10.1016/j.bpj.2016.01.016} {\bibfield
  {journal} {\bibinfo  {journal} {Biophysical Journal}\ }\textbf {\bibinfo
  {volume} {110}},\ \bibinfo {pages} {798} (\bibinfo {year}
  {2016})}\BibitemShut {NoStop}%
\bibitem [{\citenamefont {Stratton}\ \emph {et~al.}(2016)\citenamefont
  {Stratton}, \citenamefont {Warner}, \citenamefont {Wu}, \citenamefont
  {Nikolaus}, \citenamefont {Wei}, \citenamefont {Wagnon}, \citenamefont
  {Baddeley}, \citenamefont {Karatekin},\ and\ \citenamefont
  {O’Shaughnessy}}]{Stratton2016}%
  \BibitemOpen
  \bibfield  {author} {\bibinfo {author} {\bibfnamefont {B.~S.}\ \bibnamefont
  {Stratton}}, \bibinfo {author} {\bibfnamefont {J.~M.}\ \bibnamefont
  {Warner}}, \bibinfo {author} {\bibfnamefont {Z.}~\bibnamefont {Wu}}, \bibinfo
  {author} {\bibfnamefont {J.}~\bibnamefont {Nikolaus}}, \bibinfo {author}
  {\bibfnamefont {G.}~\bibnamefont {Wei}}, \bibinfo {author} {\bibfnamefont
  {E.}~\bibnamefont {Wagnon}}, \bibinfo {author} {\bibfnamefont
  {D.}~\bibnamefont {Baddeley}}, \bibinfo {author} {\bibfnamefont
  {E.}~\bibnamefont {Karatekin}},\ and\ \bibinfo {author} {\bibfnamefont
  {B.}~\bibnamefont {O’Shaughnessy}},\ }\href
  {https://doi.org/10.1016/j.bpj.2016.02.019} {\bibfield  {journal} {\bibinfo
  {journal} {Biophysical Journal}\ }\textbf {\bibinfo {volume} {110}},\
  \bibinfo {pages} {1538} (\bibinfo {year} {2016})}\BibitemShut {NoStop}%
\bibitem [{\citenamefont {Kluge}\ \emph {et~al.}(2022)\citenamefont {Kluge},
  \citenamefont {P{\"o}hnl},\ and\ \citenamefont {B{\"o}ckmann}}]{Kluge2022}%
  \BibitemOpen
  \bibfield  {author} {\bibinfo {author} {\bibfnamefont {C.}~\bibnamefont
  {Kluge}}, \bibinfo {author} {\bibfnamefont {M.}~\bibnamefont {P{\"o}hnl}},\
  and\ \bibinfo {author} {\bibfnamefont {R.~A.}\ \bibnamefont {B{\"o}ckmann}},\
  }\href {https://doi.org/10.1016/j.bpj.2022.01.029} {\bibfield  {journal}
  {\bibinfo  {journal} {Biophysical Journal}\ }\textbf {\bibinfo {volume}
  {121}},\ \bibinfo {pages} {671} (\bibinfo {year} {2022})}\BibitemShut
  {NoStop}%
\bibitem [{\citenamefont {Różycki}\ and\ \citenamefont
  {Lipowsky}(2015)}]{Lipowsky2015}%
  \BibitemOpen
  \bibfield  {author} {\bibinfo {author} {\bibfnamefont {B.}~\bibnamefont
  {Różycki}}\ and\ \bibinfo {author} {\bibfnamefont {R.}~\bibnamefont
  {Lipowsky}},\ }\href {https://doi.org/10.1063/1.4906149} {\bibfield
  {journal} {\bibinfo  {journal} {The Journal of Chemical Physics}\ }\textbf
  {\bibinfo {volume} {142}},\ \bibinfo {pages} {054101} (\bibinfo {year}
  {2015})}\BibitemShut {NoStop}%
\bibitem [{\citenamefont {Swift}\ and\ \citenamefont
  {Hohenberg}(1977)}]{Swift-Hohenberg1977}%
  \BibitemOpen
  \bibfield  {author} {\bibinfo {author} {\bibfnamefont {J.}~\bibnamefont
  {Swift}}\ and\ \bibinfo {author} {\bibfnamefont {P.~C.}\ \bibnamefont
  {Hohenberg}},\ }\href {https://doi.org/10.1103/PhysRevA.15.319} {\bibfield
  {journal} {\bibinfo  {journal} {Physical Review A}\ }\textbf {\bibinfo
  {volume} {15}},\ \bibinfo {pages} {319} (\bibinfo {year} {1977})}\BibitemShut
  {NoStop}%
\bibitem [{\citenamefont {Hoyle}(2006)}]{Hoyle2006}%
  \BibitemOpen
  \bibfield  {author} {\bibinfo {author} {\bibfnamefont {R.~B.}\ \bibnamefont
  {Hoyle}},\ }\href {https://doi.org/10.1017/CBO9780511616051} {\emph {\bibinfo
  {title} {Pattern formation: An introduction to methods}}}\ (\bibinfo
  {publisher} {Cambridge University Press},\ \bibinfo {year}
  {2006})\BibitemShut {NoStop}%
\bibitem [{\citenamefont {Shemesh}\ \emph {et~al.}(2005)\citenamefont
  {Shemesh}, \citenamefont {Geiger}, \citenamefont {Bershadsky},\ and\
  \citenamefont {Kozlov}}]{Shemesh2005}%
  \BibitemOpen
  \bibfield  {author} {\bibinfo {author} {\bibfnamefont {T.}~\bibnamefont
  {Shemesh}}, \bibinfo {author} {\bibfnamefont {B.}~\bibnamefont {Geiger}},
  \bibinfo {author} {\bibfnamefont {A.~D.}\ \bibnamefont {Bershadsky}},\ and\
  \bibinfo {author} {\bibfnamefont {M.~M.}\ \bibnamefont {Kozlov}},\ }\href
  {https://doi.org/10.1073/pnas.0500254102} {\bibfield  {journal} {\bibinfo
  {journal} {Proceedings of the National Academy of Sciences of the United
  States of America}\ }\textbf {\bibinfo {volume} {102}},\ \bibinfo {pages}
  {12383} (\bibinfo {year} {2005})}\BibitemShut {NoStop}%
\bibitem [{\citenamefont {Gov}(2006)}]{Gov2006}%
  \BibitemOpen
  \bibfield  {author} {\bibinfo {author} {\bibfnamefont {N.~S.}\ \bibnamefont
  {Gov}},\ }\href {https://doi.org/10.1529/biophysj.106.088484} {\bibfield
  {journal} {\bibinfo  {journal} {Biophysical Journal}\ }\textbf {\bibinfo
  {volume} {91}},\ \bibinfo {pages} {2844} (\bibinfo {year}
  {2006})}\BibitemShut {NoStop}%
\bibitem [{\citenamefont {Ayari}\ \emph {et~al.}(2004)\citenamefont {Ayari},
  \citenamefont {Kurdi}, \citenamefont {Vallade}, \citenamefont
  {Gulino-Debrac},\ and\ \citenamefont {Riveline}}]{Delanoe2004}%
  \BibitemOpen
  \bibfield  {author} {\bibinfo {author} {\bibfnamefont {H.~D.}\ \bibnamefont
  {Ayari}}, \bibinfo {author} {\bibfnamefont {R.~A.}\ \bibnamefont {Kurdi}},
  \bibinfo {author} {\bibfnamefont {M.}~\bibnamefont {Vallade}}, \bibinfo
  {author} {\bibfnamefont {D.}~\bibnamefont {Gulino-Debrac}},\ and\ \bibinfo
  {author} {\bibfnamefont {D.}~\bibnamefont {Riveline}},\ }\href
  {www.pnas.orgcgidoi10.1073pnas.0304297101} {\bibfield  {journal} {\bibinfo
  {journal} {Proceedings of the National Academy of Sciences of the United
  States of America}\ }\textbf {\bibinfo {volume} {101}},\ \bibinfo {pages}
  {2229} (\bibinfo {year} {2004})}\BibitemShut {NoStop}%
\end{thebibliography}

\end{document}